%% file: main.tex
\documentclass[sigconf]{acmart}

\usepackage{amsmath,amsfonts}
\usepackage{algorithm,algorithmic}
\usepackage{booktabs}
\usepackage{graphicx}
\usepackage{xcolor}
\usepackage{multirow}
\usepackage{bm}
\usepackage{subcaption}
\usepackage{colortbl}
\usepackage{ulem}
\usepackage{enumitem}

\usepackage{amssymb}
\usepackage{tcolorbox}
\usepackage{xspace}
\usepackage{cleveref}
\crefname{subsection}{§\!}{§\!§\!}
\AtBeginDocument{%
  \providecommand\BibTeX{{%
    \normalfont B\kern-0.5em{\scshape i\kern-0.25em b}\kern-0.8em\TeX}}}

\copyrightyear{2026}
\acmYear{2026}
\setcopyright{cc}
\setcctype{by}
\acmConference[CIKM '26]{Proceedings of the 35th ACM International Conference on Information and Knowledge Management}{November 07--11, 2026}{Rome, Italy}
\acmBooktitle{Proceedings of the 35th ACM International Conference on Information and Knowledge Management (CIKM '26), November 07--11, 2026, Rome, Italy}
\acmDOI{10.1145/3799682.3840906}
\acmISBN{979-8-4007-2539-5/2026/11}

\begin{document}

%%

%%
%% Custom commands
%%
\newcommand{\ours}{SCoRD}
\newcommand{\reranker}{re-ranker}
\newcommand{\Reranker}{Re-ranker}
\newcommand{\bx}{\mathbf{x}}
\newcommand{\bh}{\mathbf{h}}
\newcommand{\be}{\mathbf{e}}
\newcommand{\bq}{\mathbf{q}}
\newcommand{\bk}{\mathbf{k}}
\newcommand{\bv}{\mathbf{v}}
\newcommand{\bG}{\mathbf{G}}
\newcommand{\bH}{\mathbf{H}}
\newcommand{\bW}{\mathbf{W}}
\newcommand{\balpha}{\bm{\alpha}}
\newcommand{\btheta}{\bm{\theta}}
\newcommand{\cS}{\mathcal{S}}
\newcommand{\cD}{\mathcal{D}}
\newcommand{\cI}{\mathcal{I}}
\newcommand{\cC}{\mathcal{C}}
\newcommand{\cL}{\mathcal{L}}
\newcommand{\RR}{\mathbb{R}}
\newcommand{\proposed}{\ours{}\xspace}
\newcommand{\smallsection}[1]{{\vspace{0.03in} \noindent \bf {#1}}}

\captionsetup[table]{skip=2pt}
\captionsetup[figure]{skip=2pt}
\newlength{\textfloatsepsave} \setlength{\textfloatsepsave}{\textfloatsep} \setlength{\textfloatsep}{5pt}

\title{SCoRD: Semantic-Assisted Continual Retriever-Reranker Distillation for LLM-Based Recommendation}

\author{Seunghyun Baek}
\affiliation{
    \institution{Korea University}
    \city{Seoul}
    \country{Republic of Korea}
}
\authornote{Both authors contributed equally to this research.}
\email{seunghb320@korea.ac.kr}

\author{Gyuseok Lee}
\affiliation{
    \institution{University of Illinois at Urbana-Champaign}
    \city{Champaign}
    \state{IL}
    \country{USA}
}
\authornotemark[1]
\email{gyuseok2@illinois.edu}

\author{Seunghan Lee}
\affiliation{
    \institution{Korea University}
    \city{Seoul}
    \country{Republic of Korea}
}
\email{seunghanlee@korea.ac.kr}

\author{Wonbin Kweon}
\affiliation{
    \institution{Sungkyunkwan University}
    \city{Suwon}
    \country{Republic of Korea}
}
\email{wonbinkweon@skku.edu}

\author{Dong Wang}
\affiliation{
    \institution{University of Illinois at Urbana-Champaign}
    \city{Champaign}
    \state{IL}
    \country{USA}
}
\email{dwang24@illinois.edu}

\author{SeongKu Kang}
\affiliation{
    \institution{Korea University}
    \city{Seoul}
    \country{Republic of Korea}
}
\authornote{Corresponding author.}
\email{seongkukang@korea.ac.kr}

\renewcommand{\shortauthors}{Seunghyun Baek et al.}
%%
%% Abstract
%%
\begin{abstract}
\input{sections/000abstract}
\end{abstract}




\maketitle

%%
%% Body
%%
\input{sections/010introduction}
\input{sections/020related_work}

\section{Problem Formulation}
\input{sections/030problem_formulation}

\input{sections/040method}
\input{sections/050experiments}

\input{sections/060conclusion}

\begin{acks}
This work was supported by a Korea University Grant, ICT Creative Consilience Program through the IITP grant funded by the MSIT (IITP-2026-RS-2020-II201819), Basic Science Research Program through the NRF funded by the Ministry of Education (NRF-2021R1A6A1A03045425), the NRF grant funded by the MSIT (RS-2026-25486220), and the IITP grant funded by the MSIT (IITP-2026-RS-2026-25616664, AI Star Fellowship Support Program).
\end{acks}

\section*{GenAI Usage Disclosure}
Generative AI tool was used as part of the research methodology. In this work, LLM (LLaMA3.2-3B-Instruct) was used to infer users' underlying intents and serve as a reranking model within the proposed framework. Beyond the role in the research methodology, generative AI tools were limitedly used for writing and grammar refinement. All outputs were carefully reviewed and revised by the authors. Experimental results, datasets, and scientific contributions are entirely the authors' own work.

\bibliographystyle{ACM-Reference-Format}
\balance
\bibliography{references}

\end{document}

%% file: sections/000abstract.tex
Recommendation systems increasingly adopt a two-stage pipeline, where an ID-based retriever retrieves candidates and an LLM-based reranker refines their rankings.
To improve retrieval quality, reranker-to-retriever distillation is commonly used to transfer the reranker’s knowledge to the retriever.
For practical deployment, however, this pipeline must continually adapt to evolving interests and incoming interactions.
A naive solution is to repeatedly update the LLM reranker and distill its latest knowledge, but this incurs prohibitive costs.
Updating the retriever alone is cheaper, but its limited capacity makes adaptation from sparse data difficult.
We propose \proposed, a continual knowledge distillation framework for LLM-based reranking pipelines under a non-stationary data stream.
\proposed introduces a \textit{semantic reasoning assistant} that distills the LLM’s ability to infer underlying user intents into reusable intent-level guidance.
It selectively distills reranker knowledge to the retriever on low-confidence sequences, guides retriever-only updates without repeated LLM inference, and feeds retriever-derived representations and intent-drift signals back to the reranker.
Experiments on real-world datasets show that \proposed enables effective and efficient retriever-reranker co-adaptation.

%% file: sections/010introduction.tex
\section{Introduction}\label{sec:Introduction}
Recently, large language models (LLMs) have advanced recommendation by using world knowledge to reason over interaction sequences and capture semantic user preferences~\cite{wei2024llmrec, jia2025learn, lee2026uncertainty}.
However, directly applying LLMs on large numbers of users and items incurs prohibitive inference costs~\cite{wu2024survey, gao2023chat, zhao2024recommender}.
Recent systems therefore adopt a retrieve-then-rerank pipeline: (1) an ID-based retriever (e.g., SASRec~\cite{kang2018self}) filters the vast item space to a small candidate set, and (2) an LLM-based reranker then reorders them~\cite{luo2025recranker, hou2024large, yue2023llamarec, gao2025llm4rerank, tian2025corank}.
This balances scalability with ranking precision, enabling the practical deployment of LLMs in recommendation~\cite{yang2023palr}.

\begin{figure}[t]
  \centering
  \hspace*{-0.25cm}
  \includegraphics[width=1.02\columnwidth]{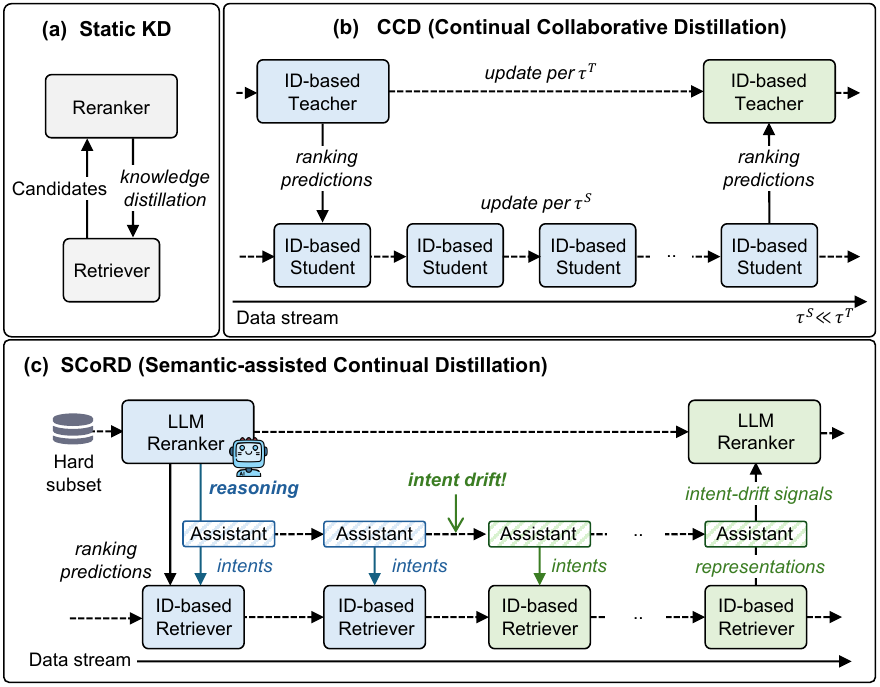}
  \caption{A conceptual comparison of (a) static distillation, (b) CCD, and (c) our approach. Best~viewed~in~color.}
  \label{fig:intro}
  \vspace{-0.1cm}
\end{figure}

As rerankers possess greater model capacity and richer ranking signals than lightweight retrievers, knowledge distillation (KD) has been widely 
employed to transfer these capabilities~\cite{cui2024distillation, jiang2023boot, li2024intermediate, wu2025bidirectional}.
Specifically, the reranker (teacher) transfers its ranking knowledge (e.g., ranked lists or ranking scores) into the retriever (student), enabling the retriever to surface high-quality candidate items (Figure~\ref{fig:intro}a).
Subsequently, these refined candidates directly improve the reranker's final performance, forming a mutually reinforcing relationship.
Consequently, compared with optimizing the two modules independently, KD improves not only retriever but also the effectiveness of the retrieve-then-rerank pipeline as a whole~\cite{ren2021rocketqav2}.

However, applying such KD under a non-stationary data stream is more challenging than one-time distillation in static settings~\cite{lee2024continual}.
In real-world applications, new users, items, and interactions continuously arrive, while user preferences and item popularity evolve over time~\cite{lee2025collaborative}.
This necessitates both the retriever and reranker to adapt accordingly.
A naive extension of KD would repeatedly update the reranker on each incoming data and distill its latest knowledge to the retriever.
However, this strategy is impractical, as it requires repeated LLM reranker updates and inference, incurring prohibitive computational overhead~\cite{hu2022lora, ouyang2022training}.\footnote{In our experiments, the LLM reranker requires 3.5--9.4$\times$ longer training time and 72.4--144.4$\times$ longer inference time than the ID-based retriever across datasets.}

In practice, the LLM reranker cannot be updated as frequently as the lightweight retriever.
This creates a dilemma.
If KD is performed without updating the reranker, its supervision becomes stale because the reranker remains tied to an older data distribution.
Consequently, the retriever receives outdated guidance and may fail to reflect recent trends or incorporate newly arrived users and items.
Alternatively, one may update the retriever alone between reranker update cycles.
Although viable, it leaves the capacity-constrained retriever to learn only from limited interaction data alone, often leading to suboptimal performance.
Thus, continual KD requires a mechanism that preserves up-to-date guidance for the retriever without repeatedly updating or invoking the LLM reranker.

One recent work~\cite{lee2024continual} addresses a related problem through continual collaborative distillation (CCD), which decouples teacher and student update cycles~(Figure \ref{fig:intro}b).
In CCD, the teacher periodically distills its ranking predictions to the student at longer intervals (e.g., weekly), while the student adapts to incoming data at shorter intervals (e.g., daily).
To mitigate student-side performance degradation, CCD uses the student’s high-ranked predictions to augment sparse interactions.
The updated student predictions are then used to augment interaction data for the next teacher update.
Thus, CCD establishes a continual KD framework over data streams, where the teacher and student exchange ranking predictions while allowing the student to adapt between teacher update cycles.

Although CCD offers a practical update scheme for continual KD, it is designed for ID-based teacher-student models.
In such settings, the teacher and student have a relatively moderate capacity gap and learn from similar collaborative signals, making direct prediction distillation effective.
However, LLM-based reranking pipelines depart from this setting, as the reranker and retriever differ fundamentally in both model capacity and knowledge source.
This raises unresolved design questions beyond the scope of CCD.

\smallsection{First, how can limited LLM supervision be used effectively? }
Unlike ID-based teachers, the LLM reranker incurs substantially higher computational costs, making uniform distillation over all incoming sequences infeasible.
Thus, LLM-based distillation should focus on sequences where the retriever most needs guidance, while providing sufficiently informative signals to enable the retriever to learn effectively from only a small subset.

\smallsection{Second, how can LLM semantic reasoning be reused during retriever-only updates? }
The LLM reranker’s predictive strength stems not merely from producing ranking scores, but from reasoning over interaction sequences to infer underlying user intents. 
However, this reasoning is accessible only when the LLM is invoked for distillation, leaving the continually updated retriever without semantic guidance until the next distillation.
Thus, this reasoning capability should be distilled into a reusable form that continuously assists the retriever without repeated LLM inference.

\smallsection{Third, how can retriever's knowledge support LLM reranker updates? }
As the retriever is updated more frequently, it captures recent collaborative patterns and evolving user interests earlier than the LLM reranker.
During reranker updates, relying solely on the LLM’s own reasoning over new interactions is inefficient and may overlook these up-to-date signals already captured by the retriever.
Thus, the reranker should use retriever-side knowledge to keep its semantic reasoning aligned with recent behavioral patterns.

As a solution, we propose \textbf{\proposed}, a \underline{S}emantic-assisted \underline{Co}ntinual \underline{R}etriever-reranker \underline{D}istillation framework~(Figure \ref{fig:intro}c).
\proposed redesigns the reranker-retriever knowledge interface with a \textit{Semantic Reasoning Assistant}, a lightweight module that distills the LLM’s reasoning ability to infer underlying user intents.
Rather than using the LLM merely as a source of ranking predictions, the assistant converts its semantic reasoning into reusable intent-level guidance.

\proposed addresses the three design questions through three stages.

\textbf{First}, during LLM-to-retriever KD, \proposed selectively applies distillation to sequences where the retriever exhibits low prediction confidence.
For these hard sequences, the reranker provides ranking predictions to the retriever, 
while also distilling LLM-inferred intents into the assistant, such as ``\textit{budget-friendly electronics}'' or ``\textit{upgrading a gaming setup}''.
The assistant then converts these intents into semantic guidance for the retriever.
Together, these prediction and intent signals provide richer supervision, allowing the retriever to learn effectively from a small LLM-supervised subset.

\textbf{Second}, during retriever-only updates, the assistant infers underlying user intents that best explain newly arriving sequences without additional LLM inference.
To improve their reliability, \proposed reinforces intent signals strongly shared across behaviorally similar sequences.
The resulting intents enhance retriever prediction and provide additional learning signals for capturing evolving user interests, helping the retriever adapt to new data.

\textbf{Third}, during reranker updates, \proposed transfers the retriever’s up-to-date behavioral knowledge back to the LLM reranker.
This knowledge is conveyed through retriever-derived representations and intent-drift signals, helping the reranker align its semantic reasoning with recent collaborative patterns.

Our contributions are summarized as follows:
\begin{itemize}[leftmargin=*] \vspace{-\topsep}

\item We highlight the underexplored challenges of updating LLM-based reranking pipelines over evolving data streams. 
To the best of our knowledge, we are the first to formulate this problem for ID-based retriever and LLM reranker co-adaptation.

\item We propose \proposed framework that facilitates continual retriever-reranker knowledge transfer through a semantic reasoning assistant, enabling efficient and effective co-adaptation.

\item We conduct extensive experiments on three real-world datasets, showing that \proposed improves both retrieval and reranking performance over state-of-the-art baselines.
\end{itemize}

%% file: sections/020related_work.tex
\section{Related Work}

\noindent
\textbf{Retriever-reranker based recommendation.}
Two-stage retrieve-then-rerank pipelines have become a practical solution for scalable recommendation~\cite{covington2016deep, huang2023cooperative}.
An ID-based retriever efficiently narrows the item space to a small candidate set, which a more expressive reranker then refines with precise scoring.
Early rerankers rely on ID-based models, scoring candidates by their feature-level relevance to user history~\cite{xi2022multi}, or jointly encoding the candidate list to capture inter-item dependencies~\cite{ai2018learning, pei2019personalized}.
More recently, LLMs have emerged as powerful rerankers by leveraging semantic reasoning over user interactions~\cite{yue2023llamarec, gao2025llm4rerank}, substantially improving ranking quality.
However, existing pipelines assume static settings, leaving their co-adaptation to an evolving data stream underexplored.

\smallsection{Knowledge distillation in recommendation.}
KD has been widely adopted to transfer knowledge from a large teacher to a compact student~\cite{tang2018ranking, kang2020rrd, kang2021topology, kang2022consensus, kang2025bpl}.
Early work focuses on ranking-based distillation, training the student to mimic the teacher's ranked outputs~\cite{lee2019collaborative, kweon2021bidirectional}. 
Subsequent work explores representation-level distillation to provide richer structural knowledge~\cite{kang2021topology, kang2023distillation}.
With the rise of LLMs, KD has been extended to leverage LLMs as teachers for lightweight models~\cite{cui2024distillation, du2025active}, and to bridge the gap between semantic and collaborative signals~\cite{sun2024large, kim2025lost}.
In retriever-reranker pipelines, KD further plays a crucial role by distilling knowledge from the reranker into the retriever, improving both retrieval quality and overall pipeline performance~\cite{ren2021rocketqav2, huang2023cooperative}.
Yet, these methods assume one-time distillation; their extension to a non-stationary data stream remains underexplored.
CCD~\cite{lee2024continual} pioneers continual KD via ranking prediction, but focuses on ID-based teacher-student models.
It does not address LLM-based reranking pipelines, where the two modules exhibit substantial asymmetry in model capacity and knowledge~source.

\smallsection{Continual learning for recommendation.}
Continual learning (CL) updates models incrementally as new data arrives~\cite{li2017learning,kirkpatrick2017overcoming}.
The goal of CL is to balance knowledge acquisition (plasticity) and retention (stability) over time~\cite{yoo2025continual, he2023dynamically}.
Existing approaches fall into two categories.
Regularization-based methods constrain model updates in the parameter space, discouraging the parameters from deviating drastically from previously learned values~\cite{wang2023structure, xu2020graphsail, wang2021graph}.
Replay-based methods store representative historical interactions and reuse them during updates~\cite{zhang2024influential, qin2025d2k, zhu2023reloop2}.
More recently, CL has been extended to LLM-based recommendation through parameter-efficient adaptation~\cite{yoo2025lora, shi2024preliminary}, such as low-rank adaptation~\cite{hu2022lora} with regularization toward its past state~\cite{yoo2025lora}.
However, existing CL studies mainly focus on single-model updates; how to integrate CL and KD for retriever-reranker co-adaptation remains an open question.

In sum, prior work has studied the above three directions largely in isolation, leaving their intersection underexplored.
This gap is practically important because the retriever and reranker must co-adapt over an evolving data stream, while repeated LLM updates and distillation incur prohibitive costs.
This work addresses this~gap.

%% file: sections/030problem_formulation.tex
\smallsection{Definition 1 (Continual Sequential Recommendation).}
Following~\cite{wang2023structure,xu2020graphsail,lee2024continual,lee2026capturing}, the data stream $\mathcal{D}$ is viewed as consecutive data blocks $[D_0, D_1, \ldots, D_k, \ldots]$.
Each block $D_k$ contains interaction sequences observed during a specific time period $\tau$. 
Here, $D_0$ serves as the base block used to build the model before continual updates on subsequent blocks.
Let $\mathcal{U}_k$ and $\mathcal{I}_k$ be the sets of users and items appearing in $D_k$, respectively.
For each user $u \in \mathcal{U}_k$, let $S^{(k)}_u = \{i_1,\ldots, i_{|S^{(k)}_u|}\}$ be the interaction sequence within $D_k$.
At $k$-th block, the system updates its model using $D_k$, without directly accessing the previous blocks $D_{<k}$.
The goal is to predict the next item over the accumulated item set given the current-block sequence, i.e., $\arg\max_{i \in \mathcal{I}_{\leq k}} P(i \mid S^{(k)}_u)$.

\smallsection{Definition 2 (Retrieval-Reranking Pipeline).}
We adopt a retrieval-reranking recommendation pipeline~\cite{hou2024large,yue2023llamarec,gao2025llm4rerank}. 
Let $M_S$ be a lightweight ID-based retriever and $M_T$ be a LLM-based reranker. 
For each user, $M_S$ first identifies a set of top-$N$ candidate items, which are then reordered by $M_T$ to produce the final recommendation list.

\smallsection{Definition 3 (Asynchronous Update Cycle of Retriever and Reranker).}
The retriever and reranker differ substantially in model size, training cost, and inference time.
Thus, the lightweight retriever can be updated frequently to reflect new interactions, whereas frequently updating the reranker is impractical.
Following the update setting of~\cite{lee2024continual}, we consider an asynchronous update cycle:
the retriever $M_S$ is updated at a short interval $\tau_S$ (e.g., daily), while the reranker $M_T$ is updated at a longer interval $\tau_T$ (e.g., weekly).
Under this cycle, the two modules must co-adapt over time despite their different update frequencies.

We define each data block based on the teacher update cycle; thus, the block interval is set to $\tau = \tau_T$.

\smallsection{Problem definition.}
We aim to enable co-adaptation of the ID-based retriever $M_S$ and the LLM reranker $M_T$ over a non-stationary data stream, while reducing prohibitive LLM update and distillation costs.
Under the asynchronous update cycle, this co-adaptation repeatedly performs three stages: (1) LLM-to-retriever distillation, (2) retriever update at $\tau_S$, and (3) reranker update at $\tau_T$.

%% file: sections/040method.tex
\section{\proposed Framework}
We first construct the semantic reasoning assistant, the core of \proposed, using the base block $D_0$ (\cref{sub:assistant}).
\proposed then uses the assistant for continual co-adaptation through three stages: reranker-to-retriever KD (\cref{sub:stage1}), retriever update (\cref{sub:stage2}), and reranker update (\cref{sub:stage3}).
Finally, we discuss the efficiency of \proposed~(\cref{sub:inference}).

\subsection{Semantic Assistant Construction on $D_0$}\label{sub:assistant}
Before continual update begins, \proposed constructs the assistant $\mathcal{A}$ on the base block $D_0$.
The purpose of $\mathcal{A}$ is to retain semantic knowledge from LLMs and provide continuous guidance to the retriever during subsequent updates.
A common approach to provide LLM guidance to an ID-based model is to use LLM-derived outputs, such as ranking results~\cite{wu2025bidirectional, cui2024distillation} or textual embeddings~\cite{liu2025improving, cui2024distillation}, as additional input features or supervision.
Although effective in static settings, this approach is impractical for continual updates over data streams, as it requires repeated LLM calls to encode newly arriving~data.

Given that the LLM’s strength lies in inferring underlying intents from interactions, we distill this capability into a compact module.
One may train a smaller language model to generate intents using LLM-inferred intents as supervision.
However, this leads to a model-size dilemma: small models may lack the generation ability to produce reliable intents, whereas larger ones undermine~efficiency.

To address this, we redesign the form of LLM reasoning knowledge from \textit{free-form generation} to \textit{memory-based selection}. 
Specifically, we organize LLM-inferred intents as a dynamic intent memory, a discrete set of intent units that expands along the data stream.
Given each sequence, the assistant selects the most relevant intents from this memory.
For example, the memory may include intent units such as “budget-friendly electronics” or “upgrading a gaming setup” from previous interactions, which can later be used to comprehend newly arriving sequences. 
This design provides continuous semantic guidance without frequent LLM calls, while reducing the hallucination risks from free-form generation.

\begin{figure}[t]
  \centering
  \hspace*{-0.25cm}
  \includegraphics[width=1\columnwidth]{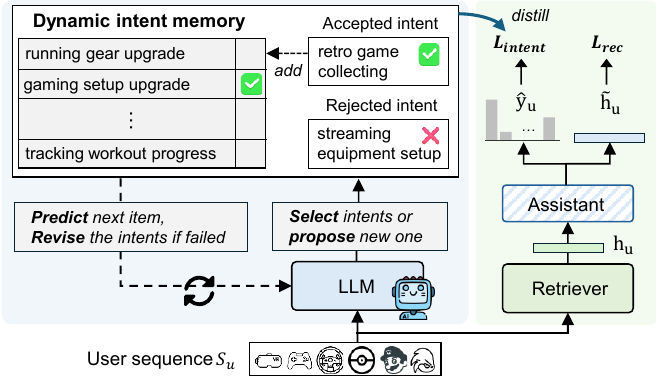}
%  \vspace{-3mm}
  \caption{Semantic reasoning assistant construction process.}
  \label{fig:assist}
  \vspace{-0.1cm}
\end{figure}

Below, we describe memory construction and assistant training on the base block $D_0$.
The overview is presented in Figure~\ref{fig:assist}.

\subsubsection{\textbf{Dynamic intent memory construction}.}\label{sec:4.1.1}
Let $\mathcal{G}$ be a dynamic intent memory that incrementally expands along the data stream, initialized as $\mathcal{G}^{(0)}=\emptyset$.
For each user sequence $S_u^{(0)}$ in the base block $D_0$, we construct the memory through three steps: (1) intent inference, (2) intent verification, and (3) memory update.

\smallsection{Intent inference.}
We instruct the LLM to infer intents by \textit{selecting} suitable ones from the current memory or \textit{proposing} new intents when no existing intent sufficiently explains the sequence:
\begin{equation}\label{eq:intent_inference}
\hat{\mathcal{G}}_u = \mathrm{LLM}\bigl((S_u^{(0)}, \mathcal{G}^{(0)}) ; \mathcal{P}_{\mathrm{intent}}\bigr),
\end{equation}
where $\hat{\mathcal{G}}_u$ denotes the candidate intents for user $u$, and $\mathcal{P}_{\mathrm{intent}}$ is the corresponding prompt.\footnote{The key instruction is: ``Given $S_u^{(0)}[:-1]$ and $\mathcal{G}^{(0)}$, select the intents that best explain the user's underlying preferences. If no suitable intent exists, propose new intents''.}

\smallsection{Intent verification.}
To obtain reliable intents, we evaluate whether the inferred intents can explain the user’s observed behavior.
Following the verification task of~\cite{liu2026diagnostic, lee2025sprint}, we hold out the last item of $S_u$ and ask the LLM to identify it among $N\!+\!1$ candidates (one held-out item and $N$ random negatives), given the remaining sequence and $\hat{\mathcal{G}}_u$. We set $N\!=\!5$.

If the LLM selects the correct item, the inferred intents are accepted.
Otherwise, we provide the failed prediction as feedback and ask the LLM to revise the intents. 
This repeats until the verification succeeds, up to three trials.
If it still fails, we do not record the intent for user $u$; the user may be reconsidered in a future block~(\cref{subsub:hard}), after more interactions have~accumulated.

\smallsection{Memory update.}
Finally, the verified intents are incorporated into the dynamic memory:
\begin{equation}\label{eq:memory_update}
\mathcal{G}^{(0)} \leftarrow \mathcal{G}^{(0)} \cup \mathcal{G}_u,
\end{equation}
where $\mathcal{G}_u$ denotes the verified intents for user $u$.
Existing intents are reused, while newly identified intents are added to expand the memory.
Thus, $\mathcal{G}^{(0)}$ grows only when the current memory lacks suitable intents for explaining a user sequence, preventing unnecessary expansion while covering diverse interactions.

As a result, we obtain the initial dynamic intent memory $\mathcal{G}^{(0)}$ and the verified intent assignments $\mathcal{G}_u$ for users in the base block.

\subsubsection{\textbf{Joint training of assistant and retriever.}}
Given the initial memory $\mathcal{G}^{(0)}$ and the intent assignments $\{\mathcal{G}_u\}_{u\in \mathcal{U}_0}$, we train the assistant $\mathcal{A}$ together with the retriever $M_S$.
The goal is to distill the LLM’s intent-inference capability into $\mathcal{A}$, so that it can predict the underlying intents and provide semantic guidance to the retriever. 
As a sequence may have multiple intents, we formulate this intent prediction as a multi-label classification over $\mathcal{G}^{(0)}$.

\smallsection{Memory-based intent prediction.}
For each user sequence $S_u$, the retriever produces a sequence representation $\mathbf{h}_u = M_S(S_u) \in \mathbb{R}^d$.\footnote{For sequential retrievers (e.g., SASRec~\cite{kang2018self}), $\mathbf{h}_u$ is the hidden state at the last position.}
To predict relevant intents from the memory, $\mathcal{A}$ adopts a query-key-value design~\cite{vaswani2017attention}, where $\mathbf{h}_u$ serves as the query and each intent $c \in \mathcal{G}^{(0)}$ is represented by a learnable embedding $\mathbf{p}_c \in \mathbb{R}^d$:
\begin{equation}
\mathbf{q}_u = \mathbf{W}_q \mathbf{h}_u, \quad
\mathbf{k}_c = \mathbf{W}_k \mathbf{p}_c, \quad
\mathbf{v}_c = \mathbf{W}_v \mathbf{p}_c,
\end{equation}
where $\mathbf{W}_q, \mathbf{W}_k, \mathbf{W}_v \in \mathbb{R}^{d \times d}$ are learnable projection matrices.
The user-intent relevance is computed as: $\hat{y}_{u,c}=\sigma(\mathbf{q}_u^\top\mathbf{k}_c)$, where $\sigma$ is the sigmoid function.

\smallsection{Semantic-guided representation.}
While the ID-based retriever captures collaborative patterns, it lacks explicit semantic knowledge of user intents.
Using the intent relevance scores, $\mathcal{A}$ converts the intents into semantic guidance for the retriever. 
Specifically, it aggregates the values of relevant intents and injects them into the original sequence representation:
\begin{equation}\label{eq:h_tilde_u}
    \tilde{\mathbf{h}}_u = \text{LayerNorm}\bigg(\mathbf{h}_u + \text{Dropout}\bigg(\sum_{c \in \mathcal{G}^{(0)}} \phi(\hat{y}_{u,c})\mathbf{v}_c\bigg)\bigg),
\end{equation}
where $\phi(n)\!=\!n \cdot \mathbb{I}[n > x]$ retains scores above threshold~$x$. We set $x\!=\!0.5$ in this work.
This produces a semantic-guided representation $\tilde{\mathbf{h}}_u$ by focusing on the intents most relevant to the current sequence.

\smallsection{Optimization.}
We jointly optimize the assistant $\mathcal{A}$, parameterized by $\{\mathbf{W}_q, \mathbf{W}_k, \mathbf{W}_v, \{\mathbf{p}_c\}_{c \in \mathcal{G}}\}$, and the retriever $M_S$ with two objectives: reasoning distillation and semantic-guided recommendation.
\begin{enumerate}[leftmargin=*] \vspace{-0.5\topsep}
    \item \textbf{Reasoning distillation}: 
    This aligns the predicted user-intent relevance scores with the LLM-inferred intents.
    Given $\mathcal{G}_u$, we define the binary label $y_{u,c} = \mathbb{I}[c \in \mathcal{G}_u]$ and optimize:
    \begin{equation}\label{eq:L_intent_origin}
        \hspace{-0.2cm}
        \mathcal{L}_{\text{intent}} = -\frac{1}{|\mathcal{G}^{(0)}|}\sum_{c \in \mathcal{G}^{(0)}} \bigg[ y_{u,c} \log \hat{y}_{u,c} + (1 - y_{u,c}) \log(1 - \hat{y}_{u,c}) \bigg].
    \end{equation}

    \item \textbf{Semantic-guided recommendation}:
    This aligns the semantic-guided representation $\tilde{\mathbf{h}}_u$ with 
    the next-item prediction task:
    \begin{equation}\label{eq:L_rec}
        \hspace{-0.2cm}
        \mathcal{L}_{\text{rec}} = -\sum_{S_u \in D_0} \bigg[ \log \sigma(\mathbf{e}_{i_{\text{pos}}}^\top \tilde{\mathbf{h}}_u) + \sum_{i_{\text{neg}} \in \mathcal{N}_u^-} \log(1 - \sigma(\mathbf{e}_{i_{\text{neg}}}^\top \tilde{\mathbf{h}}_u)) \bigg],
    \end{equation}
    where $\mathbf{e}_i$ is the retriever item embedding, $i_{\text{pos}}$ is the ground-truth next item, and $\mathcal{N}_u^-$ is the negative item set.
\end{enumerate}
The joint training objective is:
\begin{equation}
\mathcal{L}_{\mathcal{A}, M_S} =  \mathcal{L}_{\text{rec}} + \lambda_{\text{intent}} \mathcal{L}_{\text{intent}},
\end{equation}
where $\lambda_{\text{intent}}$ is the loss-balancing hyperparameter.
This jointly trains $\mathcal{A}$ to predict LLM-derived intents and $M_S$ to use the resulting semantic guidance for next-item prediction.
The reranker $M_T$ is also warmed up on $D_0$ using the standard cross-entropy loss.

The following sections describe how $\mathcal{A}$ supports continual distillation. 
Assume that the retriever and reranker have been trained up to block $D_{k-1}$.
Each new cycle starts with distilling the recent reranker knowledge to the retriever.
The distilled retriever then serves as the starting point for subsequent updates on $D_k$.

\begin{figure*}[t]
    \centering    
    \includegraphics[width=0.9\textwidth]{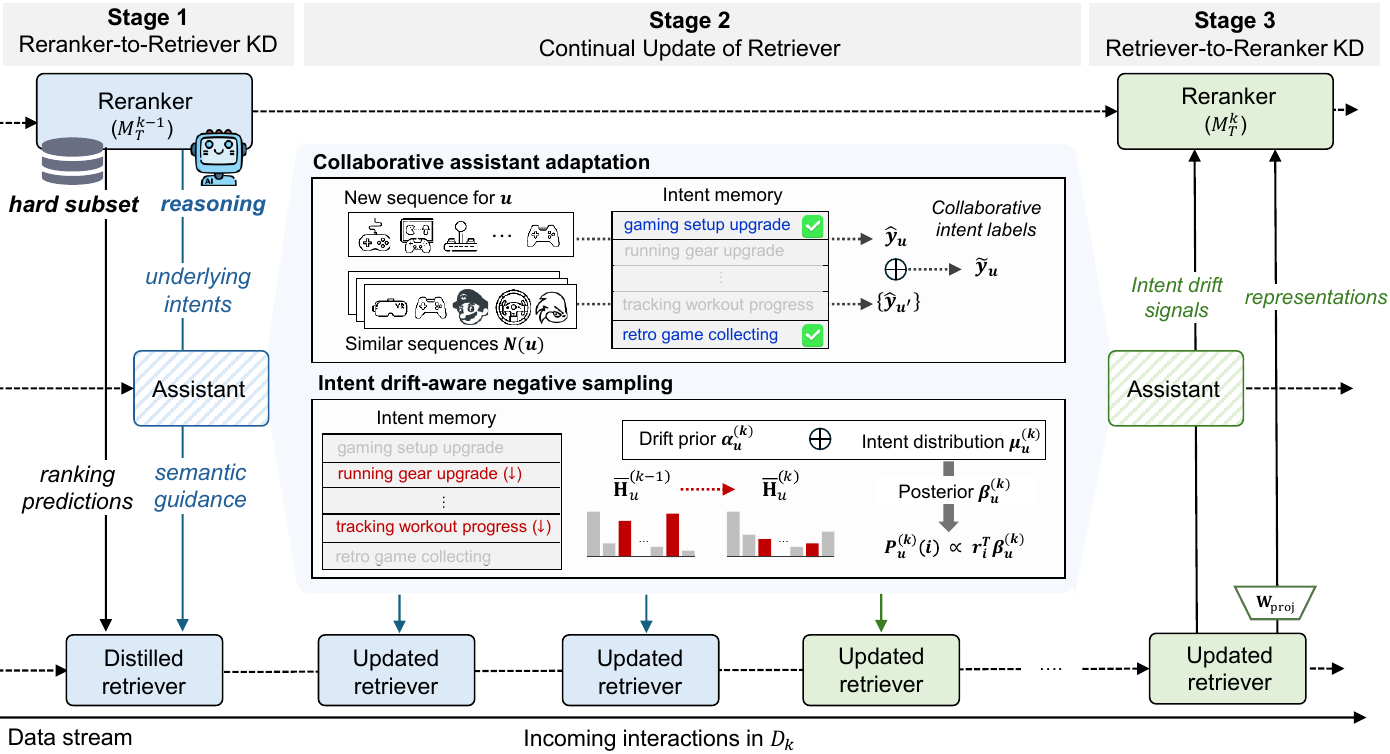}
    \caption{Overview of \proposed framework. Best viewed in color.
    }
    \label{fig:method}
    \vspace{-0.5cm}
\end{figure*}

\subsection{Stage 1: Reranker-to-Retriever KD}\label{sub:stage1}
In this stage, we use the reranker trained up to $D_{k-1}$ to update the retriever and assistant before processing $D_k$, so that they start the new block with recent reranker knowledge.
We first identify hard sequences based on the retriever’s confidence, so that LLM guidance is concentrated where it is most needed. 
For these sequences, we distill two forms of supervision from the LLM reranker: fine-grained ranking signals to the retriever and inferred intents to the assistant.

\subsubsection{\textbf{Confidence-based hard sequence selection.}}\label{subsub:hard}
Instead of distilling over all sequences, we prioritize sequences where the retriever has low prediction confidence. 
We estimate this confidence from sequence-item alignment: a well-encoded sequence representation should be closely aligned to the items in the sequence.
For each sequence $S_u \in D_{k-1}$, we compute the confidence score:
\begin{equation}
    \rho_u = \frac{1}{|S_{u}|} \sum_{i \in S_{u}} \sigma(\mathbf{e}_{i}^\top \tilde{\mathbf{h}}_u).
\end{equation}
A lower $\rho_u$ indicates that the retriever cannot confidently account for the in-sequence items.
We select the bottom $r\%$ of sequences based on $\rho_u$ to form $D^{\text{hard}}_{k-1}$, thereby reducing the cost of distillation.
This avoids redundant LLM distillation on sequences that the retriever already predicts well. 
Semantic knowledge from these hard sequences is later propagated to broader sequences in Stage~2~(\cref{subsub:col_sem}).

\subsubsection{\textbf{Semantic-assisted ranking distillation.}}
Given the selected hard sequences $D^{\text{hard}}_{k-1}$, we first update the intent memory and then perform ranking distillation with semantic guidance from~$\mathcal{A}$.

\smallsection{Intent memory update.}
For each hard sequence, we update the previous memory $\mathcal{G}^{(k-1)}$ using the same inference-verification-update procedure in~\cref{sec:4.1.1}.
This yields the updated memory $\mathcal{G}^{(k)}$ and verified intent assignments $\mathcal{G}_u$ for users in the hard sequence set, with new intents added when needed.

\smallsection{Ranking distillation with semantic guidance.}
We adopt the listwise ranking distillation loss used in CCD~\cite{lee2024continual}, which trains the student to imitate the teacher’s ranking order.
Our distinction is that the distillation loss is computed with the semantic-guided representation $\tilde{\mathbf{h}}_u$.
This allows $\mathcal{A}$ to be jointly optimized with the retriever while guiding ranking transfer with intent semantics.

Let $\pi_u$ denote the item ranking list produced by the reranker for user $u$, and let $\pi_u(j)$ be the $j$-th item in the list. 
We define the top-$N$ permutation probability~\cite{xia2008listwise} using the retriever’s prediction scores and minimize its negative log-likelihood:
\begin{equation}
\mathcal{L}_{\mathrm{KD}} =
-\sum_{S_u \in D^{\mathrm{hard}}_{k-1}} \log \prod_{j=1}^{N} \frac{\exp(\tilde{s}_{u,\pi_u(j)})}{
\sum_{i=j}^{|\pi_u|} \exp(\tilde{s}_{u,\pi_u(i)}) },
\end{equation}
where $\tilde{s}_{u,i} = \tilde{\mathbf{h}}_u^\top \mathbf{e}_i$ is the semantic-guided item prediction score of the retriever.

\subsubsection{\textbf{Optimization of stage 1}}
The overall objective is defined~as:
\begin{equation}\label{eq:L_stage1}
    \min_{{\theta}_S, \theta_\mathcal{A}} \mathcal{L}_{\text{stage1}} = \mathcal{L}_{\text{KD}} + \lambda_{\text{intent}} \mathcal{L}_{\text{intent}},
\end{equation}
where ${\theta}_S$ and $\theta_{\mathcal{A}}$ are the parameters of the retriever and assistant, respectively.
$\mathcal{L}_{\text{intent}}$ is computed as in Eq.~\eqref{eq:L_intent_origin} using $\mathcal{G}^{(k)}$ and the verified intents for $D^{\mathrm{hard}}_{k-1}$.

\subsection{Stage 2: Continual Update of Retriever}\label{sub:stage2}
In this stage, the retriever $M_S^{(k)}$ is continuously updated on the incoming block $D_k$ every $\tau_S$.
The key challenge is enabling the capacity-constrained retriever to effectively incorporate new sequences and adapt to evolving interests. 
The assistant $\mathcal{A}$ is designed to provide continuous semantic guidance: given a sequence, it infers the underlying intents and supports semantic-guided prediction. 
However, $\mathcal{A}$ must also adapt, since new sequences lack verified intent labels and user interests may drift over time. 
We propose two new techniques to provide up-to-date semantic guidance.

\subsubsection{\textbf{Collaborative semantic assistant adaptation.}}\label{subsub:col_sem}
$\mathcal{A}$ should accurately predict intents for sequences that have not been directly supervised with intent labels.
To this end, we draw inspiration from self-training \cite{selftrain1, selftrain2}, a well-established semi-supervised learning technique that improves generalization by using pseudo-labels on unlabeled data.
However, directly using $\mathcal{A}$'s intent predictions as pseudo-intent labels can be unstable and prone to noise.
Instead, we propose to incorporate intents from behaviorally similar sequences.

Specifically, for each user $u$, we retrieve the top-$K$ behaviorally similar user sequences and construct the collaborative intent label $\tilde{y}_{u,c}$ by aggregating their predicted intent scores:
\begin{equation}
\tilde{y}_{u,c} = \frac{1}{2}\bigg(\hat{y}_{u,c} \,+ \sum_{u' \in \mathcal{N}(u)} w_{u,u'} \cdot \hat{y}_{u',c}\bigg),
\end{equation}
where $\mathcal{N}(u) = \text{TopK}(\{\cos(\mathbf{h}_u, \mathbf{h}_{u'})\}_{u' \in \mathcal{B}})$ denotes the top-$K$ similar users in mini-batch $\mathcal{B}$, and $w_{u,u'}$ is the softmax-normalized cosine similarity over $\mathcal{N}(u)$.
For newly arriving sequences, we compute the collaborative intent learning loss $\mathcal{L}_{\text{co-intent}}$ by replacing $y_{u,c}$ with $\tilde{y}_{u,c}$ in Eq.~\eqref{eq:L_intent_origin}.
This reinforces intent signals that are strongly and consistently shared across behaviorally similar sequences, providing more reliable semantic guidance.

\subsubsection{\textbf{Intent drift-aware negative sampling.}}\label{subsub:sem_neg}
As user interests evolve over time, some intents fade while new ones emerge.
To capture such drift, negative sampling is critical: negatives drawn from items aligned with \textit{faded intents} provide more effective training signals for adapting to evolving interests than random negatives.
To this end, we quantify how the user's intent distribution shifts across blocks and use this shift to guide negative sampling.

\smallsection{Intent histogram construction.}
We use the intent distribution as a semantic proxy for the user's interests.
To estimate the current intent distribution, we construct a user intent histogram $\mathbf{H}_u^{(k)}$ by aggregating item-level intent relevance over the sequence $S_u^{(k)}$:
\begin{equation}
    \mathbf{H}_u^{(k)} = \sum_{i \in S_u^{(k)}} \mathbf{r}_i, \quad
    \mathbf{r}_i = [\sigma((\mathbf{W}_q \mathbf{e}_i)^\top (\mathbf{W}_k \mathbf{p}_c))]_{c \in \mathcal{G}^{(k)}},
\end{equation}
where $\mathbf{r}_i \in \mathbb{R}^{|\mathcal{G}^{(k)}|}$ denotes the relevance vector between item $i$ and intents in $\mathcal{G}^{(k)}$.
We normalize it as $\overline{\mathbf{H}}_u^{(k)} = \mathbf{H}_u^{(k)} /\; \|\mathbf{H}_u^{(k)}\|_1$ and compare it across consecutive blocks to identify faded intents.

\smallsection{Faded-intent negative sampling.}
To identify faded intents, we first compute a faded-intent score $\boldsymbol{\alpha}_u^{(k)} \in \mathbb{R}^{|\mathcal{G}^{(k)}|}$ by comparing consecutive intent histograms.
Intents whose normalized frequency decreases from block $k-1$ to $k$ receive higher weights\footnote{$\,\overline{\mathbf{H}}_u^{(k-1)}$ is zero-padded over newly introduced intents in $\mathcal{G}^{(k)}$.}:
\begin{equation}
\boldsymbol{\alpha}_u^{(k)} = \text{softmax}\left(\overline{\mathbf{H}}_u^{(k-1)} - \overline{\mathbf{H}}_u^{(k)}\right).
\end{equation}
While this score captures intent drift, using it directly for negative sampling can be unstable because it is estimated only from sparse observed interactions. 
A decrease in an intent may reflect either a true fading interest or insufficient observations in the current~block.

To obtain a more robust distribution, we treat $\boldsymbol{\alpha}_u^{(k)}$ as a drift prior and refine it with evidence from plausible unobserved items. 
Specifically, we consider the top-$Q$ unobserved items $\mathcal{Q}_u$\footnote{We set $Q{=}100$ in our experiments.} from the retriever and compute their mean intent relevance: $\boldsymbol{\mu}_u^{(k)} = \frac{1}{Q}\sum_{i \in \mathcal{Q}_u} \mathbf{r}_i$.
This reflects the intent distribution of retriever-plausible items.
%This captures the intent distribution of items that the retriever considers plausible. 
Inspired by Dirichlet-Multinomial conjugate updates~\cite{bishop2006pattern}, we combine the prior with this evidence to obtain a smoothed~posterior:
\begin{equation}
    \boldsymbol{\beta}_u^{(k)} = (\boldsymbol{\alpha}_u^{(k)} + \boldsymbol{\mu}_u^{(k)}) / \,\,{\|\boldsymbol{\alpha}_u^{(k)} + \boldsymbol{\mu}_u^{(k)}\|_1}.
\end{equation}
We then sample negatives from top-$Q$ unobserved items with probability proportional to their alignment with the posterior:
$P_u^{(k)}(i) \propto \mathbf{r}_i^\top \boldsymbol{\beta}_u^{(k)}$, yielding intent-drift $\mathcal{N}_u^-$ for $\mathcal{L}_{\text{rec}}$ in Eq.~\eqref{eq:L_rec}.
This probability becomes high when item $i$ is strongly associated with the faded intents emphasized by $\boldsymbol{\beta}_u^{(k)}$, providing semantic drift-aware negatives that help the retriever adapt to evolving user preferences.

\subsubsection{\textbf{Optimization of stage 2}}
The overall objective is:
\begin{equation}\label{eq:L_stage2}
\min_{\theta_S, \theta_\mathcal{A}} \mathcal{L}_{\text{stage2}} = \mathcal{L}_{\text{rec}} + \lambda_{\text{intent}} \mathcal{L}_{\text{co-intent}}  + \lambda_{\text{reg}} \mathcal{L}_{\text{reg}},
\end{equation}
where $\mathcal{L}_{\text{rec}}$ uses the intent drift-aware negatives, and $\mathcal{L}_{\text{co-intent}}$ uses collaborative pseudo-intent labels.
We additionally adopt a regularization term widely used in continual learning~\cite{yoo2025embracing, xu2020graphsail, wang2021graph, wang2023structure}, 
which penalizes excessive shifts in representations to prevent catastrophic forgetting: $\mathcal{L}_{\text{reg}} = \sum_{u \in \mathcal{U}_{k-1} \cap \mathcal{U}_k} \|\mathbf{h}_u^{(k-1)}\!- \mathbf{h}_u^{(k)}\|_2^2$.

\subsection{Stage 3: Retriever-to-Reranker KD}\label{sub:stage3}
Once the cycle $\tau_T$ is due, the reranker is updated on $D_k$.
For effective adaptation, we leverage up-to-date retriever knowledge to reduce the knowledge gap caused by the reranker's infrequent updates.

\smallsection{Retriever-side knowledge injection.}
Following~\cite{li2023e4srec}, we incorporate item embeddings from the ID-based retriever into prompt construction. 
Specifically, each item embedding $\mathbf{e}_i$ from the retriever is projected into the LLM input embedding space as $\mathbf{x}_i = \mathbf{W}_{\text{proj}}\mathbf{e}_i$.
The interaction history is then represented by the projected embeddings in the encoding prompt for recommendation.\footnote{The encoding prompt is: ``Given the interaction history $[\mathbf{x}_{i_1}, \dots, \mathbf{x}_{i_n}]$, extract representations for next-item recommendation.''}

The LLM reranker then produces a sequence representation from this prompt. 
We attach an item prediction layer~\cite{li2023e4srec}, implemented as a linear layer with softmax output.
Let $\mathbf{p}_u^{(k)}$ denote the item prediction distribution given $S_u^{(k)}$, where $\mathbf{p}_u^{(k)}[i]$ is the predicted probability of item $i$.
We train the reranker with the standard cross-entropy loss for next-item prediction:
\begin{equation}
\mathcal{L}_{\mathrm{CE}} = -\sum_{S_u^{(k)} \in D_k} \log \mathbf{p}_u^{(k)}[i_{\mathrm{pos}}],
\end{equation}
where $i_{\mathrm{pos}}$ is the ground-truth next item.

\smallsection{Drift-aware negative signals.}
To explicitly reflect intent drift captured by $\mathcal{A}$, we introduce an auxiliary cross-entropy loss $\mathcal{L}_{\mathrm{drift}}$ that compares the ground-truth next item $i_{\mathrm{pos}}$ against intent-drift negatives $\mathcal{N}_u^-$.
This is implemented by normalizing the softmax only over $i_{\mathrm{pos}}$ and $\mathcal{N}_u^-$, providing focused supervision against faded-intent items and complementing the standard full-item prediction~loss.

\subsubsection{\textbf{Optimization of stage 3}}
The overall objective is:
\begin{equation}\label{eq:L_stage3}
    \min_{{\theta}_T, \mathbf{W}_{\text{proj}}}  \mathcal{L}_{\text{stage3}} = \mathcal{L}_{\text{CE}} + \lambda_\text{drift} \mathcal{L}_{\text{drift}},
\end{equation}
where $\theta_T$ denotes the trainable parameters of the reranker, comprising the LoRA adapters~\cite{hu2022lora} and item prediction layer.
$\lambda_{\mathrm{drift}}$ controls the strength of the drift-aware auxiliary loss.

\subsection{Training and Inference Efficiency}\label{sub:inference}

\smallsection{Training.}
\proposed addresses the computational bottleneck in continual KD: 
repeated LLM inference, distillation, and fine-tuning.
\begin{itemize}[leftmargin=*] \vspace{-\topsep}
    \item \textbf{Stage 1.} LLM distillation is performed only on the top-$r\%$ hardest sequences in each block; we set $r=20$ in our experiments.
    \item \textbf{Stage 2.} Frequent retriever updates are assisted by semantic guidance from $\mathcal{A}$, requiring no additional LLM calls.
    \item \textbf{Stage 3.} Reranker fine-tuning is performed only every $\tau_T$, avoiding updates at every retriever cycle.
\end{itemize} \vspace{-\topsep}
Together, these designs reduce LLM inference and update costs while maintaining continuous semantic guidance, making retriever--reranker co-adaptation over a non-stationary data stream practical.
Figure~\ref{fig:method} and Algorithm 1 present the overall procedure.

\smallsection{Inference.}
Given $S_u$, the retriever ranks all items using the semantic-guided representation $\tilde{\mathbf{h}}_u$, where relevant intents are selected by $\mathcal{A}$.
Items are scored by $\tilde{\mathbf{h}}_u^\top \mathbf{e}_i$, yielding top-$N$ candidates for reranking.\footnote{We set $N{=}20$ in our experiments.}
The reranker then scores each candidate using the predicted distribution $\mathbf{p}_u$.
Intent prediction adds negligible overhead to the total latency; we provide a detailed analysis in Section~\ref{result:eff}.

\input{tables/algorithm}

%% file: tables/algorithm.tex
\begin{algorithm}[t]
\small
\caption{SCoRD algorithm over data blocks}
\label{alg:scord}
\begin{algorithmic}[1]

\STATE \textbf{Input:}~$M_S^{(k-1)}$, $M_T^{(k-1)}$, $\mathcal{A}$ with intent memory $\mathcal{G}^{(k-1)}$
\STATE \textbf{Output:}~Updated $M_S^{(k)}$, $M_T^{(k)}$, $\mathcal{A}$, $\mathcal{G}^{(k)}$

\vspace{3pt}
\STATE \textit{/* Base Block */}
\IF{$k = 0$}
  \STATE Train $M_T^{(0)}$ on ${D}_0$
  \FOR{each sequence $S_u^{(0)} \in {D}_0$}
    \STATE Infer \& verify intent $\rightarrow$ update memory $\mathcal{G}^{(0)}$ \hfill $\triangleright$ Eqs.\,(1--2)
  \ENDFOR
  \STATE Jointly train $\mathcal{A}$ and $M_S^{(0)}$ on ${D}_0$ \hfill $\triangleright$ Eq.\,(7)
  \STATE \textbf{return}
\ENDIF

\vspace{2pt}
\STATE \textit{/* Incremental Block */}
\STATE \textbf{Stage 1: Reranker-to-retriever KD}
\STATE Construct ${D}^{\mathrm{hard}}$ and update $\mathcal{G}^{(k-1)} \!\to\! \mathcal{G}^{(k)}$ %\hfill $\triangleright$ 
\STATE Update $M_S$ and $\mathcal{A}$ \hfill $\triangleright$ Eq.\,(10)

\vspace{2pt}
\STATE \textbf{Stage 2: Continual retriever update}
\FOR{every $\tau_S$}
  \STATE Enhance intent labels and sample intent-drift $\mathcal{N}_u^-$ \hfill $\triangleright$ Eqs.\,(11--14)
  \STATE Update $M_S$ and $\mathcal{A}$ \hfill $\triangleright$ Eq.\,(15)
\ENDFOR

\vspace{2pt}
\STATE \textbf{Stage 3: Retriever-to-reranker KD}
\STATE Project item embeddings and sample intent-drift $\mathcal{N}_u^-$
\STATE Update $M_T$ \hfill $\triangleright$ Eq.\,(17)
\end{algorithmic}
\end{algorithm}

%% file: sections/050experiments.tex
\section{Experiments}\label{sec:experiments}
\subsection{Experimental Setup}\label{sub:exp_setup}
\input{tables/data_statistics_table}
\input{tables/main_final}
%% ---------------------------------------------------------
\noindent
\textbf{Datasets.}
We use three real-world datasets: Books (Amazon)~\cite{mcauley2015image}, Yelp~\cite{yelp_open_dataset}, and Movies \& TV (Amazon)~\cite{mcauley2015image}, using the most recent 3, 5, and 2 years of interactions, respectively.
Following existing CL setups~\cite{lee2026capturing, wang2023structure,xu2020graphsail}, we simulate a non-stationary data stream by chronologically splitting each dataset into five interaction blocks: the base block $D_0$ contains the earliest 60\% of interactions for initial training, while the remaining 40\% is equally divided into four incremental blocks $D_1$--$D_4$ (10\% each) for continual updates.
For each block, we apply $k$-core filtering~\cite{seidman1983network} with a user threshold of 5 for all datasets and item thresholds of 3, 3, and 2 for Books, Yelp, and Movies \& TV, respectively.
Each block consists of interaction sequences, where the last item serves as the test label, the second-to-last as the validation label, and the remaining items for training.
Table~\ref{tab:data_statistics} presents detailed statistics per block for each dataset, including the number of accumulated intents (i.e., $|\mathcal{G}^{(k)}|$).

%% ---------------------------------------------------------
\smallsection{Evaluation protocol.}
Following prior CL studies~\cite{wang2023structure, mi2020ader}, models are trained on $D_0$ and continually updated on each incremental block without access to previous data.
For each block, we evaluate both the reranker and the retriever using Hit Rate (H)~\cite{Hit} and NDCG (N)~\cite{NDCG} at cutoffs \{5, 10, 20\}, following~\cite{wu2025bidirectional, lee2024continual, liu2025improving, yoo2025embracing}.
Following~\cite{lee2026capturing}, we report results on $D_2$--$D_4$, as preference drift is less pronounced in the early stages.
While some CL studies~\cite{do2023continual, lee2024continual} adopt retained average ($\frac{1}{t}\sum_{i=1}^{t} a_{t,i}$, where $a_{t,i}$ is the performance on $D_i$ after training on $D_t$) as a metric, recent work emphasizes plasticity in adapting to newly arriving data~\cite{yoo2025embracing, yoo2025continual}. Following this trend, we report the performance of the reranker and retriever at each~block,
averaged over five independent runs with different random seeds.
%% ---------------------------------------------------------

\smallsection{Retriever-Reranker setup.}
We simulate a retriever-reranker pipeline.
For a fair comparison, all methods use the same backbone configuration: E4SRec~\cite{li2023e4srec} (reranker), which injects ID-based item embeddings into an LLM for reranking, and SASRec~\cite{kang2018self} (retriever), which captures sequential preferences via self-attention~\cite{vaswani2017attention}.
Note that the reranker and retriever are paired for each method.

%% ---------------------------------------------------------
\smallsection{Baselines.}
We group baselines into four categories and compare \proposed with state-of-the art methods in a reranker–retriever pipeline.

\begin{enumerate}[leftmargin=*]\vspace{-\topsep}
    \item \textbf{Reference methods}: serve as upper and lower bounds.
        \begin{itemize}[leftmargin=*]\vspace{-\topsep}
            \item \textbf{Full-Batch} retrains from scratch on all accumulated data. 
            
            \item \textbf{Fine-Tune} updates models only on the current data block.
        \end{itemize}
    
    \item \textbf{CL-based methods}: employ regularization or replay for CL.
        \begin{itemize}[leftmargin=*]\vspace{-\topsep}
            \item \textbf{PISA}~\cite{yoo2025embracing} is a regularization-based method that balances stability and plasticity per user based on preference shifts.
            
            \item \textbf{Reloop2}~\cite{zhu2023reloop2} is a replay-based method that uses error memory to self-correct future recommendations.
            
        \end{itemize}

    \item \textbf{Continual KD}: applies KD across incremental data blocks.
        \begin{itemize}[leftmargin=*]\vspace{-\topsep}
            \item \textbf{CCD}~\cite{lee2024continual} pioneers continual KD via ranking predictions~between teacher-student models with asynchronous updates.
        \end{itemize}
    
    \item \textbf{LLM-based KD methods}: transfer knowledge between LLM-based rerankers and ID-based retrievers.
        \begin{itemize}[leftmargin=*]\vspace{-\topsep}
            \item \textbf{LLMD4Rec}~\cite{wu2025bidirectional} jointly refines reranker and retriever via bidirectional exchange of semantic and collaborative~signals.
        
            \item \textbf{CoT-Rec}~\cite{liu2025improving} uses Chain-of-Thought (CoT)~\cite{wei2022chain} reasoning to generate text descriptions, encodes them as semantic embeddings, and initializes the embeddings of the retriever.
        \end{itemize}
\end{enumerate}\vspace{-\topsep}
As LLMD4Rec and CoT-Rec assume a static environment, we extend them to the CL setting: both methods update the retriever every $\tau_s$; at every $\tau_T$, LLMD4Rec performs bidirectional KD while CoT-Rec provides the reranker with the updated retriever's item embeddings.

\smallsection{Implementation details.}
Hyperparameters are tuned via grid~search on the validation set.
Baseline hyperparameters follow the search ranges in the original papers.
We describe details in four~aspects. 
\begin{itemize}[leftmargin=*]\vspace{-\topsep}
    \item \textbf{Common.}
    All models are implemented in PyTorch and trained on NVIDIA A100 GPUs using Adam~\cite{kingma2015adam}.
    Following CCD~\cite{lee2024continual}, we set the retriever update interval $\tau_S$ to one-tenth of the reranker update interval $\tau_T$.
    For intent generation, each item is represented by its title, category, and five sampled reviews, following~\cite{ren2024representation, liu2025enhancing}.

    \item \textbf{Reranker.}
    We fine-tune Llama-3.2-3B-Instruct~\cite{meta2024llama32} with LoRA~\cite{hu2022lora} (rank $r{=}16$, $\alpha{=}16$, dropout${=}0.1$) for 4 epochs.
    The learning rate is set to $3\text{e-}4$ and $L_2$ regularization to $5\text{e-}2$.

    \item \textbf{Retriever.}
    We use SASRec~\cite{kang2018self} ($d{=}128$, 2 layers, 2 heads) trained for 100 epochs.
    We search the learning rate over $\{1\text{e-}3, 1\text{e-}2\}$ and $L_2$ regularization over $\{5\text{e-}5, 1\text{e-}4\}$.
    We use two negatives.
   
    \item \textbf{\proposed hyperparameters.}
    We search $\lambda_{\text{intent}} \in \{0.5, 1.0, 1.5\}$, $\lambda_{\text{reg}} \in \{0.3, 0.5, 0.7\}$, and $\lambda_{\text{drift}} \in \{0.1, 0.2, 0.5\}$.
    The number of negatives for $\mathcal{L}_{\text{drift}}$ is searched over $\{5, 10\}$.
    
\end{itemize}

\vspace{-0.2cm}

\subsection{Performance Comparison}

\subsubsection{\textbf{Main results}}

Table~\ref{tab:main_table} presents the performance comparison.
Overall, \proposed outperforms all baselines across all datasets and metrics.
We analyze these gains from three perspectives.

\smallsection{Overall.}
We attribute the overall gains of \proposed to two key advantages.
First, \proposed surpasses CL-based methods (PISA, Reloop2) by enabling mutual knowledge transfer over time, whereas CL methods update each model independently.
Second, \proposed gains over CCD by leveraging auxiliary semantic guidance from the assistant beyond surface-level ranking predictions.

\smallsection{Reranker side.}
\proposed beats LLM-based KD methods (LLMD4Rec, CoT-Rec).
This is attributed to two external guidance sources:
(1) collaborative signals from the retriever and (2) intent drift-aware negatives from the assistant, which other methods do not exploit.

\smallsection{Retriever side.}
\proposed surpasses LLM-based KD methods as the assistant provides continuous guidance during retriever-only updates via semantic-guided representation and negatives.~However, other methods offer no guidance until the LLM is updated again.

Interestingly, \proposed yields stronger gains on fine-grained metrics (N@5 and H@5) than on H@20. 
We conjecture that continuous semantic assistance mainly enhances fine-grained recommendation, while adaptive control of assistant guidance could further improve broader candidate coverage. 
We leave this direction for future work.

%% ---------------------------------------------------------

\begin{figure}[t]
  \centering
  \includegraphics[width=1.0\columnwidth]{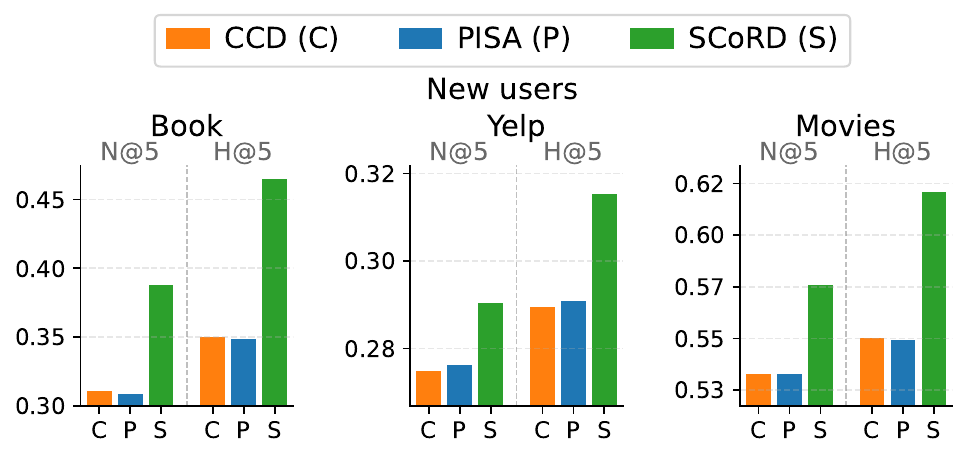}
  \vspace{-0.2cm}
  \includegraphics[width=1.0\columnwidth]{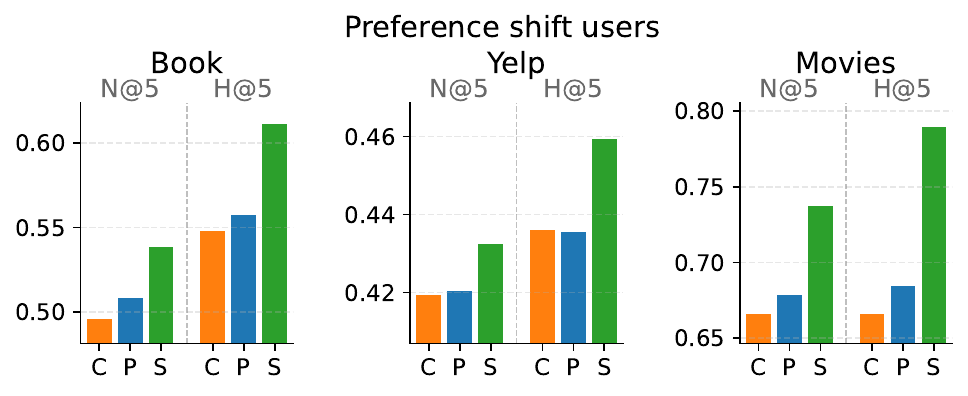}
\caption{Performance on new and preference shift users.}
  \label{fig:user_analysis}
  \vspace{-0.1cm}
\end{figure}

\subsubsection{\textbf{Analysis under CL setup.}}
We evaluate reranker performance of \proposed against CCD and PISA under the CL setup, focusing on challenging user groups and stability-plasticity balance.

\smallsection{Analysis on user groups.}
Figure~\ref{fig:user_analysis} reports performance (N@5, H@5) on new users and preference shift users\footnote{Preference shift users are the top 25\% by cosine distance $1\;-\;\cos(\bar{\mathbf{e}}^{(k-1)}_u, \bar{\mathbf{e}}^{(k)}_u)$, where $\bar{\mathbf{e}}^{(k)}_u$ is the mean item embedding given $S^{(k)}_u$ from the fine-tuned SASRec backbone.} across three datasets.
\proposed consistently outperforms CCD and PISA across all setups.
The analysis for each group is as follows.
For new users, CCD and PISA perform at a similar level, suggesting that CL methods struggle to handle users with no historical interactions.
\proposed addresses this through intent-level guidance from $\mathcal{A}$, which provides semantic context even in the absence of prior user history.
For preference shift users, PISA slightly outperforms CCD, as PISA is explicitly designed to handle preference shifts.
However, \proposed achieves substantially larger gains over both, showing that intent-drift aware learning effectively captures evolving user preferences.

Together, these results indicate that \proposed is effective not only in overall recommendation quality but also in handling the two most challenging user groups under a non-stationary data stream.

\input{tables/sp_balance}
\smallsection{Stability-plasticity analysis.}
Following~\cite{lee2024continual, lee2026capturing, do2023continual}, we adopt two CL-specific metrics\footnote{RA $= \frac{1}{t}\sum_{i=1}^{t} a_{t,i}$ (for stability); LA $= \frac{1}{t}\sum_{i=1}^{t} a_{i,i}$ (for plasticity), where $a_{ij}$ is performance on block $j$ after training on block $i$. We report results on $D_2$ through $D_4$.}: 
Retained average (RA) for stability and Learning average (LA) for plasticity, both computed on N@5.
We also report their harmonic mean (H-mean) to summarize their balance.

Table~\ref{tab:sp_balance} shows that \proposed improves over both baselines on RA, LA, and H-mean simultaneously, confirming that stability and plasticity are achieved without trading one off against the other.
We attribute the stability gain to item embedding regularization in the retriever, whose stable representations are projected into the reranker, enabling more stable user representation construction.
The plasticity gain is driven by intent-drift aware guidance from the assistant and collaborative signals from the retriever, which facilitate adaptation to newly arriving data.

\input{tables/acc_eff}
\subsubsection{\textbf{Accuracy-efficiency trade-off analysis.}}\label{result:eff}
Table~\ref{tab:acc_eff} presents accuracy (N@5, H@5) and time for semantic intent generation (Gen), model training (Tr.), and inference (Inf.) over $D_2$--$D_4$.
\proposed achieves the best performance while maintaining competitive efficiency.
Specifically, \proposed reduces generation time\footnote{LLMD4Rec reports no generation time (N/A) as it relies solely on prediction logits without generating any semantic content.} by ${\sim}10\times$ compared to CoT-Rec (e.g., 1.4h vs. 15.4h on Book).
For training, LLMD4Rec takes the longest due to full-sequence distillation,
CoT-Rec the shortest as it fine-tunes each model independently without cross-model distillation, and \proposed falls in between.
Inference time is comparable across all~methods. %, confirming no additional overhead at deployment.
As a result, \proposed achieves a better balance between recommendation quality and efficiency.

\input{tables/ablation_avg}
\subsection{Study of \proposed}
\subsubsection{\textbf{Ablation study.}}
Table~\ref{tab:ablation_avg} reports reranker and retriever performance averaged over $D_2$--$D_4$ to examine the contribution of each component on the Book dataset.
Overall, the full method with all components achieves the best performance, confirming the contribution of each. 
We analyze the results stage by stage as follows.

\smallsection{Stage 1.} Ablating $\mathcal{A}$ removes both $\mathcal{L}_{\text{intent}}$ supervision and intent-guided user representation $\tilde{\mathbf{h}}_u$, causing the largest drop.
Removing $\mathcal{L}_{\text{KD}}$ alone leads to a smaller drop, while removing both $\mathcal{A}\;\&\;\mathcal{L}_{\text{KD}}$ yields the worst performance, showing their complementary roles.

\smallsection{Stage 2.} Replacing intent-drift negatives with random negatives causes the largest drop, highlighting their importance for learning evolving user interests.
Replacing the semantic-guided representation $\tilde{\mathbf{h}}_u$ with the original retriever representation $\mathbf{h}_u$ leads to the second largest drop, showing the importance of semantic-guidance during retriever updates.
Removing $\mathcal{L}_{\text{co-intent}}$ and $\mathcal{L}_{\text{reg}}$ leads to relatively small drops compared to other components.

\smallsection{Stage 3.} Removing $\mathcal{L}_{\text{drift}}$ degrades reranker performance, indicating the importance of drift-aware training signals.
Note that retriever-side knowledge injection cannot be ablated, as item embeddings from the retriever are integral to the LLM prompt construction.

\subsubsection{\textbf{Hyperparameter analysis}}
We analyze key hyperparameters for each stage (N@5, H@5), reporting retriever performance for stages 1–2 and reranker for stage 3.

\begin{figure}[t]
  \includegraphics[width=1.0\columnwidth]{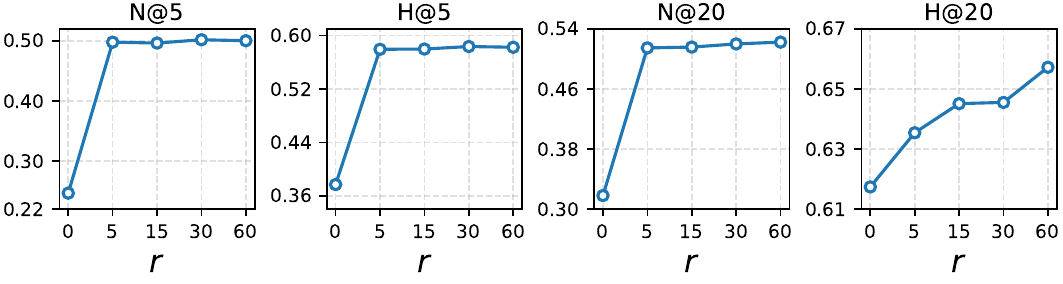}
    \caption{Effect of hard sequences selection ratio $r$.}
    \label{fig:hard}
  \vspace{-0.3cm}
\end{figure}

\smallsection{Stage 1.}
Figure~\ref{fig:hard} shows the effect of the hard sequences selection ratio $r$.
Performance improves sharply from $r{=}0$ to $r{=}5$ and saturates thereafter for most metrics.
This implies that targeting a small proportion of hard users is sufficient for meaningful gains without requiring LLM supervision over the full user set.

\begin{figure}[t]
  \includegraphics[width=0.8\columnwidth]{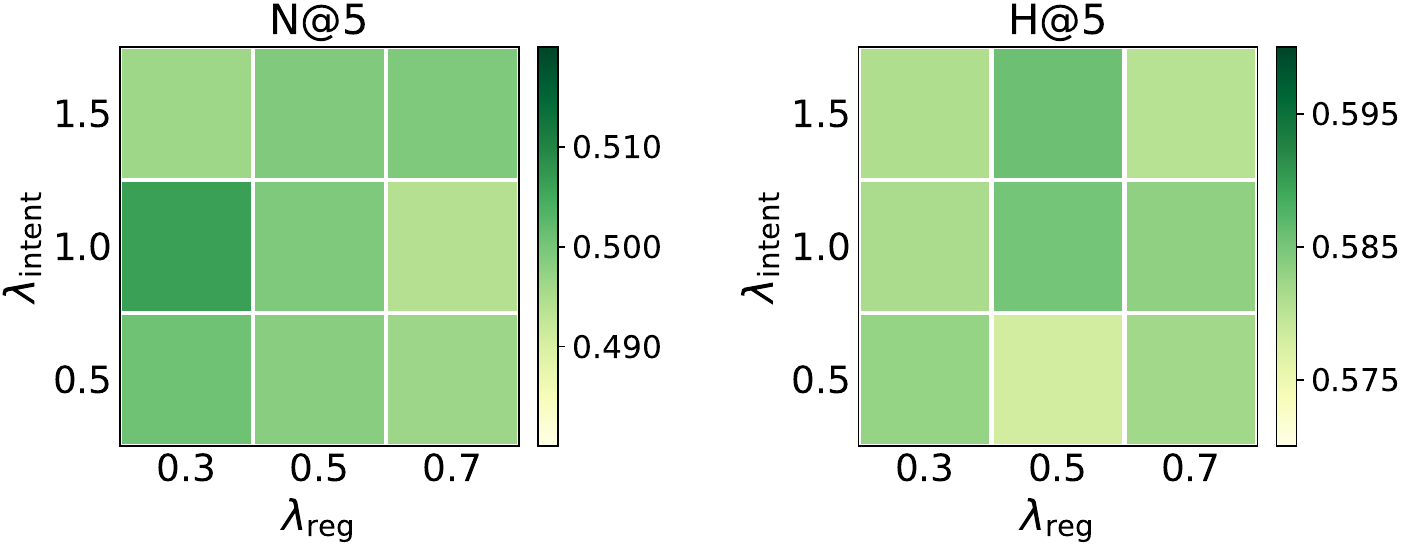}
    \caption{Effect of $\lambda_{\text{intent}}$ and $\lambda_{\text{reg}}$ on retriever performance.}
  \label{fig:retriever}
  \vspace{-0.1cm}
\end{figure}

\smallsection{Stage 2.}
Figure~\ref{fig:retriever} examines how the loss weights $\lambda_{\text{intent}}$ and $\lambda_{\text{reg}}$ jointly affect retriever updates.
Overall, $\lambda_{\text{intent}}{=}1.0$ yields the best results, with performance remaining stable across $\lambda_{\text{reg}} \in \{0.3, 0.5\}$, suggesting that \proposed is not overly sensitive to loss weight selection within the search range.

\begin{figure}[t]
  \includegraphics[width=1.0\columnwidth]{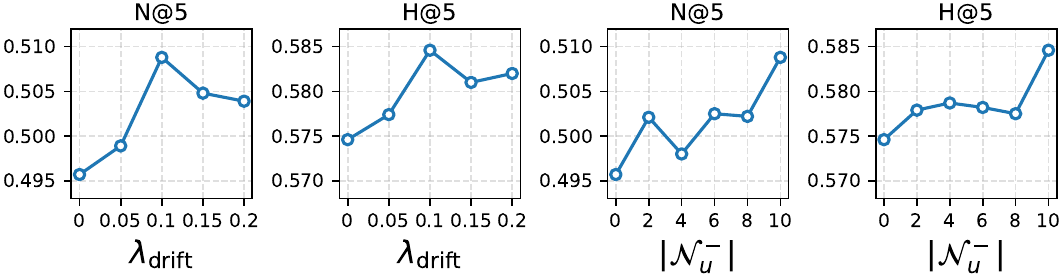}
\caption{Effect of $\lambda_{\text{drift}}$ and $|\mathcal{N}_u^-|$ on reranker performance.}  \label{fig:reranker}
  \vspace{-0.1cm}
\end{figure}

\smallsection{Stage 3.}
Figure~\ref{fig:reranker} presents the effect of $\lambda_{\text{drift}}$ and the number of intent-drift negatives $|\mathcal{N}_u^-|$ on reranker training.
For $\lambda_{\text{drift}}$, performance peaks at $0.1$ and slightly degrades beyond that, suggesting that too strong a drift signal can hurt recommendation accuracy. 
For $|\mathcal{N}_u^-|$, performance consistently improves as more intent-drift negatives are added, indicating that a larger set of faded-intent negatives provides richer contrastive supervision.
In both cases, removing these components (i.e., setting them to zero) leads to notable performance drops, validating the effectiveness of explicit intent drift-aware supervision.

%% file: tables/data_statistics_table.tex
\begin{table}[t]
\caption{Data block statistics after preprocessing.}
\centering
\renewcommand{\arraystretch}{0.9}
\renewcommand{\tabcolsep}{1.1mm}
\resizebox{\columnwidth}{!}{%
\begin{tabular}{cc|c|cccc}
\toprule
\multicolumn{2}{c|}{\textbf{Data Blocks}} & $\mathbf{D_0}$ \textbf{(60\%)} & $\mathbf{D_1}$ \textbf{(10\%)} & $\mathbf{D_2}$ \textbf{(10\%)} & $\mathbf{D_3}$ \textbf{(10\%)} & $\mathbf{D_4}$ \textbf{(10\%)} \\ \midrule\midrule
\multicolumn{1}{c|}{} & \textbf{\#  users (new)} & 30,031(30,031) & 1,529(218) & 1,735(376) & 2,021(510) & 1,822(524) \\
\multicolumn{1}{c|}{} & \textbf{\#  items (new)} & 38,981(38,981) & 2,051(1,599) & 2,241(1,854) & 2,649(2,116) & 2,664(2,278) \\
\multicolumn{1}{c|}{} & \textbf{\#  interactions} & 358,600 & 12,693 & 14,957 & 17,673 & 16,644 \\
\multicolumn{1}{c|}{} & \textbf{\#   intents} & 140 & 154 & 167 & 180 & 192 \\
\multicolumn{1}{c|}{} & \textbf{Avg. Seq Length} & 11.94 & 8.30 & 8.62 & 8.74 & 9.14 \\
\multicolumn{1}{c|}{\multirow{-6}{*}{\rotatebox[origin=c]{90}{\textbf{Book}}}} & \textbf{Sparsity} & 0.9997 & 0.9960 & 0.9962 & 0.9967 & 0.9966 \\ \hline
\multicolumn{1}{c|}{} & \textbf{\#  users (new)} & 61,734(61,734) & 1,270(463) & 1,998(815) & 3,005(1,420) & 2,527(1,030) \\
\multicolumn{1}{c|}{} & \textbf{\#  items (new)} & 47,616(47,616) & 1,982(144) & 3,061(424) & 4,407(608) & 3,820(408) \\
\multicolumn{1}{c|}{} & \textbf{\#  interactions} & 744,167 & 10,184 & 16,169 & 25,688 & 21,421 \\
\multicolumn{1}{c|}{} & \textbf{\#   intents} & 72 & 78 & 80 & 83 & 88 \\
\multicolumn{1}{c|}{} & \textbf{Avg. Seq Length} & 12.05 & 8.02 & 8.09 & 8.55 & 8.48 \\
\multicolumn{1}{c|}{\multirow{-6}{*}{\rotatebox[origin=c]{90}{\textbf{Yelp}}}} & \textbf{Sparsity} & 0.9997 & 0.9960 & 0.9974 & 0.9981 & 0.9978 \\ \hline
\multicolumn{1}{c|}{} & \textbf{\#  users (new)} & 20,380(20,380) & 700(520) & 955(751) & 598(449) & 627(459) \\
\multicolumn{1}{c|}{} & \textbf{\#  items (new)} & 37,591(37,591) & 1,851(274) & 2,425(390) & 1,481(299) & 1,494(297) \\
\multicolumn{1}{c|}{} & \textbf{\#  interactions} & 197,918 & 5,564 & 7,775 & 4,705 & 4,714 \\
\multicolumn{1}{c|}{} & \textbf{\#   intents} & 61 & 63 & 64 & 65 & 66 \\
\multicolumn{1}{c|}{} & \textbf{Avg. Seq Length} & 9.71 & 7.95 & 8.14 & 7.87 & 7.52 \\
\multicolumn{1}{c|}{\multirow{-6}{*}{\rotatebox[origin=c]{90}{\textbf{Movies}}}} & \textbf{Sparsity} & 0.9997 & 0.9957 & 0.9966 & 0.9947 & 0.9950 \\ \bottomrule
\end{tabular}
}
\vspace{-0.1cm}
\label{tab:data_statistics}
\end{table}

%% file: tables/main_final.tex
% \usepackage{multirow}

% statistical significance (p<0.05) for the t-test against the best baseline.
% $Imp$ denotes the improvement of \proposed over the best baseline.
\begin{table*}[ht!]
\caption{The overall performance comparison.
* denotes $p < 0.05$ for the paired t-test on \proposed with the best baseline.}
%\vspace{-4mm}
\label{tab:main_table}
\footnotesize
\renewcommand{\arraystretch}{0.88}
\renewcommand{\tabcolsep}{1.2mm}
\centering
\resizebox{\linewidth}{!}{
\begin{tabular}{c|l|cccccc|cccccc|cccccc}
\hline
 \multicolumn{2}{c|}{\multirow{2}{*}{\textbf{Book}}} & \multicolumn{6}{c|}{\textbf{After $D_2$}} & \multicolumn{6}{c|}{\textbf{After $D_3$}} & \multicolumn{6}{c}{\textbf{After $D_4$}} \\
\multicolumn{2}{c|}{} & \textbf{N@5} & \textbf{H@5} & \textbf{N@10} & \textbf{H@10} & \textbf{N@20} & \textbf{H@20} & \textbf{N@5} & \textbf{H@5} & \textbf{N@10} & \textbf{H@10} & \textbf{N@20} & \textbf{H@20} & \textbf{N@5} & \textbf{H@5} & \textbf{N@10} & \textbf{H@10} & \textbf{N@20} & \textbf{H@20} \\
\hline\hline
\multirow{8}{*}{\rotatebox[origin=c]{90}{\textbf{Reranker}}}
& Full-Batch
& 0.4579 & 0.5177 & 0.4716 & 0.5603 & 0.4910 & 0.6383
& 0.3612 & 0.4436 & 0.3824 & 0.5086 & 0.3992 & 0.5754
& 0.3293 & 0.4122 & 0.3567 & 0.4967 & 0.3774 & 0.5792 \\
& Fine-Tune
& 0.4185 & 0.4638 & 0.4311 & 0.5026 & 0.4490 & 0.5746
& 0.4392 & 0.4933 & 0.4523 & 0.5343 & 0.4682 & 0.5983
& 0.4972 & 0.5416 & 0.5110 & 0.5850 & 0.5301 & 0.6623 \\
& PISA
& 0.4465 & \underline{0.4956} & 0.4632 & \underline{0.5473} & 0.4816 & \textbf{0.6208}
& 0.4502 & \underline{0.5085} & 0.4658 & \underline{0.5570} & 0.4853 & \textbf{0.6351}
& 0.4981 & 0.5505 & 0.5138 & 0.5992 & 0.5318 & 0.6715 \\
& Reloop2
& 0.4261 & 0.4756 & 0.4426 & 0.5270 & 0.4588 & 0.5916
& 0.4249 & 0.4852 & 0.4363 & 0.5202 & 0.4502 & 0.5761
& 0.4913 & 0.5517 & 0.5058 & 0.5968 & 0.5212 & 0.6594 \\
& CCD
& 0.4477 & 0.5000 & 0.4623 & 0.5458 & 0.4807 & 0.6197
& 0.4324 & 0.4987 & 0.4475 & 0.5456 & 0.4657 & 0.6195
& 0.5072 & 0.5567 & 0.5218 & \underline{0.6024} & 0.5403 & \underline{0.6765} \\
& LLMD4Rec
& \underline{0.4515} & 0.4904 & \underline{0.4677} & 0.5410 & \underline{0.4846} & 0.6094
& 0.4456 & 0.4783 & 0.4598 & 0.5223 & 0.4751 & 0.5845
& \underline{0.5249} & 0.5514 & \underline{0.5393} & 0.5956 & \underline{0.5531} & 0.6517 \\
& CoT-Rec
& 0.4359 & 0.4752 & 0.4526 & 0.5273 & 0.4711 & 0.6020
& \underline{0.4535} & 0.5004 & \underline{0.4698} & 0.5510 & \underline{0.4879} & 0.6243
& 0.5153 & \underline{0.5640} & 0.5283 & 0.6048 & 0.5444 & 0.6700 \\
& \cellcolor[HTML]{D9D9D9}SCoRD
& \cellcolor[HTML]{D9D9D9}\textbf{0.4925*}
& \cellcolor[HTML]{D9D9D9}\textbf{0.5606*}
& \cellcolor[HTML]{D9D9D9}\textbf{0.5055*}
& \cellcolor[HTML]{D9D9D9}\textbf{0.6009*}
& \cellcolor[HTML]{D9D9D9}\textbf{0.5099*}
& \cellcolor[HTML]{D9D9D9}\underline{0.6179}
& \cellcolor[HTML]{D9D9D9}\textbf{0.4953*}
& \cellcolor[HTML]{D9D9D9}\textbf{0.5761*}
& \cellcolor[HTML]{D9D9D9}\textbf{0.5078*}
& \cellcolor[HTML]{D9D9D9}\textbf{0.6141*}
& \cellcolor[HTML]{D9D9D9}\textbf{0.5115*}
& \cellcolor[HTML]{D9D9D9}\underline{0.6282}
& \cellcolor[HTML]{D9D9D9}\textbf{0.5386*}
& \cellcolor[HTML]{D9D9D9}\textbf{0.6172*}
& \cellcolor[HTML]{D9D9D9}\textbf{0.5538*}
& \cellcolor[HTML]{D9D9D9}\textbf{0.6635*}
& \cellcolor[HTML]{D9D9D9}\textbf{0.5579*}
& \cellcolor[HTML]{D9D9D9}\textbf{0.6795} \\
\hline
\multirow{8}{*}{\rotatebox[origin=c]{90}{\textbf{Retriever}}}
& Full-Batch
& 0.2424 & 0.3666 & 0.2893 & 0.5109 & 0.3218 & 0.6383
& 0.1942 & 0.3035 & 0.2391 & 0.4431 & 0.2726 & 0.5754
& 0.1853 & 0.2835 & 0.2328 & 0.4299 & 0.2706 & 0.5792 \\
& Fine-Tune
& 0.1927 & 0.2967 & 0.2404 & 0.4449 & 0.2734 & 0.5746
& 0.2292 & 0.3539 & 0.2745 & 0.4927 & 0.3015 & 0.5983
& 0.3128 & 0.4463 & 0.3551 & 0.5764 & 0.3771 & 0.6623 \\
& PISA
& \underline{0.3244} & \underline{0.4531} & \underline{0.3550} & \underline{0.5477} & \underline{0.3737} & \textbf{0.6208}
& \underline{0.3003} & \underline{0.4310} & \underline{0.3394} & \underline{0.5504} & \underline{0.3608} & \textbf{0.6351}
& 0.3487 & 0.4817 & 0.3851 & 0.5945 & 0.4047 & 0.6715 \\
& Reloop2
& 0.1565 & 0.2627 & 0.2144 & 0.4424 & 0.2526 & 0.5916
& 0.1775 & 0.2893 & 0.2302 & 0.4538 & 0.2614 & 0.5761
& 0.3051 & 0.4537 & 0.3472 & 0.5823 & 0.3669 & 0.6594 \\
& CCD
& 0.3158 & 0.4372 & 0.3492 & 0.5388 & 0.3698 & \underline{0.6197}
& 0.2944 & 0.4215 & 0.3306 & 0.5328 & 0.3527 & 0.6195
& \underline{0.3787} & \underline{0.5171} & \underline{0.4111} & \underline{0.6163} & \underline{0.4264} & \underline{0.6765} \\
& LLMD4Rec
& 0.1914 & 0.3171 & 0.2433 & 0.4775 & 0.2770 & 0.6094
& 0.2115 & 0.3276 & 0.2570 & 0.4681 & 0.2866 & 0.5845
& 0.3201 & 0.4637 & 0.3559 & 0.5732 & 0.3759 & 0.6517 \\
& CoT-Rec
& 0.2223 & 0.3407 & 0.2673 & 0.4797 & 0.2985 & 0.6020
& 0.2427 & 0.3703 & 0.2876 & 0.5085 & 0.3171 & 0.6243
& 0.2880 & 0.4327 & 0.3339 & 0.5729 & 0.3587 & 0.6700 \\
& \cellcolor[HTML]{D9D9D9}SCoRD
& \cellcolor[HTML]{D9D9D9}\textbf{0.4787*}
& \cellcolor[HTML]{D9D9D9}\textbf{0.5606*}
& \cellcolor[HTML]{D9D9D9}\textbf{0.4895*}
& \cellcolor[HTML]{D9D9D9}\textbf{0.5942*}
& \cellcolor[HTML]{D9D9D9}\textbf{0.4956*}
& \cellcolor[HTML]{D9D9D9}0.6179
& \cellcolor[HTML]{D9D9D9}\textbf{0.4866*}
& \cellcolor[HTML]{D9D9D9}\textbf{0.5740*}
& \cellcolor[HTML]{D9D9D9}\textbf{0.4986*}
& \cellcolor[HTML]{D9D9D9}\textbf{0.6108*}
& \cellcolor[HTML]{D9D9D9}\textbf{0.5031*}
& \cellcolor[HTML]{D9D9D9}\underline{0.6282}
& \cellcolor[HTML]{D9D9D9}\textbf{0.5327*}
& \cellcolor[HTML]{D9D9D9}\textbf{0.6210*}
& \cellcolor[HTML]{D9D9D9}\textbf{0.5442*}
& \cellcolor[HTML]{D9D9D9}\textbf{0.6561*}
& \cellcolor[HTML]{D9D9D9}\textbf{0.5501*}
& \cellcolor[HTML]{D9D9D9}\textbf{0.6795} \\

 \hline \hline
 \multicolumn{2}{c|}{\multirow{2}{*}{\textbf{Yelp}}} & \multicolumn{6}{c|}{\textbf{After $D_2$}} & \multicolumn{6}{c|}{\textbf{After $D_3$}} & \multicolumn{6}{c}{\textbf{After $D_4$}} \\
\multicolumn{2}{c|}{} & \textbf{N@5} & \textbf{H@5} & \textbf{N@10} & \textbf{H@10} & \textbf{N@20} & \textbf{H@20} & \textbf{N@5} & \textbf{H@5} & \textbf{N@10} & \textbf{H@10} & \textbf{N@20} & \textbf{H@20} & \textbf{N@5} & \textbf{H@5} & \textbf{N@10} & \textbf{H@10} & \textbf{N@20} & \textbf{H@20} \\
\hline\hline
\multirow{8}{*}{\rotatebox[origin=c]{90}{\textbf{Reranker}}}
& Full-Batch
& 0.3953 & 0.4119 & 0.4028 & 0.4350 & 0.4098 & 0.4631
& 0.3633 & 0.4178 & 0.3744 & 0.4518 & 0.3823 & 0.4827
& 0.2461 & 0.3025 & 0.2656 & 0.3630 & 0.2816 & 0.4264 \\
& Fine-Tune
& 0.3381 & 0.3505 & 0.3436 & 0.3677 & 0.3527 & 0.4038
& 0.4083 & \underline{0.4297} & 0.4138 & \underline{0.4472} & 0.4199 & 0.4716
& 0.3758 & 0.3935 & 0.3817 & 0.4118 & 0.3906 & 0.4473 \\
& PISA
& 0.3261 & 0.3402 & 0.3325 & 0.3599 & 0.3421 & 0.3983
& 0.3949 & 0.4158 & 0.3999 & 0.4311 & 0.4077 & 0.4624
& 0.3692 & 0.3844 & 0.3761 & 0.4061 & 0.3864 & 0.4476 \\
& Reloop2
& 0.3342 & 0.3512 & 0.3416 & 0.3741 & 0.3490 & 0.4035
& 0.3885 & 0.4210 & 0.3942 & 0.4387 & 0.4013 & 0.4669
& 0.3672 & 0.4000 & 0.3733 & 0.4190 & 0.3818 & \underline{0.4534} \\
& CCD
& 0.3235 & 0.3425 & 0.3296 & 0.3615 & 0.3394 & 0.4006
& 0.3988 & 0.4185 & 0.4041 & 0.4351 & 0.4118 & 0.4660
& 0.3731 & 0.3865 & 0.3794 & 0.4064 & 0.3884 & 0.4424 \\
& LLMD4Rec
& 0.3409 & 0.3522 & \underline{0.3479} & 0.3741 & 0.3539 & 0.3980
& \underline{0.4145} & 0.4273 & \underline{0.4196} & 0.4432 & \textbf{0.4268} & 0.4720
& \underline{0.3859} & 0.3979 & \underline{0.3916} & 0.4157 & \underline{0.3982} & 0.4419 \\
& CoT-Rec
& \underline{0.3413} & \underline{0.3567} & 0.3478 & \underline{0.3770} & \underline{0.3574} & \underline{0.4154}
& 0.3869 & 0.4230 & 0.3944 & 0.4460 & 0.4028 & \textbf{0.4798}
& 0.3664 & \underline{0.4005} & 0.3734 & \underline{0.4223} & 0.3820 & \textbf{0.4572} \\
& \cellcolor[HTML]{D9D9D9}SCoRD
& \cellcolor[HTML]{D9D9D9}\textbf{0.3505*}
& \cellcolor[HTML]{D9D9D9}\textbf{0.3748*}
& \cellcolor[HTML]{D9D9D9}\textbf{0.3568*}
& \cellcolor[HTML]{D9D9D9}\textbf{0.3941*}
& \cellcolor[HTML]{D9D9D9}\textbf{0.3642*}
& \cellcolor[HTML]{D9D9D9}\textbf{0.4235*}
& \cellcolor[HTML]{D9D9D9}\textbf{0.4150}
& \cellcolor[HTML]{D9D9D9}\textbf{0.4456*}
& \cellcolor[HTML]{D9D9D9}\textbf{0.4202}
& \cellcolor[HTML]{D9D9D9}\textbf{0.4617*}
& \cellcolor[HTML]{D9D9D9}\underline{0.4237}
& \cellcolor[HTML]{D9D9D9}\underline{0.4756}
& \cellcolor[HTML]{D9D9D9}\textbf{0.3942}
& \cellcolor[HTML]{D9D9D9}\textbf{0.4204*}
& \cellcolor[HTML]{D9D9D9}\textbf{0.3982}
& \cellcolor[HTML]{D9D9D9}\textbf{0.4328*}
& \cellcolor[HTML]{D9D9D9}\textbf{0.4008}
& \cellcolor[HTML]{D9D9D9}0.4431 \\
\hline
\multirow{8}{*}{\rotatebox[origin=c]{90}{\textbf{Retriever}}}
& Full-Batch
& 0.2233 & 0.3358 & 0.2509 & 0.4197 & 0.2619 & 0.4631
& 0.1676 & 0.2744 & 0.2063 & 0.3934 & 0.2289 & 0.4827
& 0.1203 & 0.1964 & 0.1575 & 0.3119 & 0.1866 & 0.4264 \\
& Fine-Tune
& 0.1881 & 0.2786 & 0.2123 & 0.3528 & 0.2253 & 0.4038
& 0.2427 & 0.3494 & 0.2671 & 0.4241 & 0.2792 & 0.4716
& 0.2169 & 0.3160 & 0.2418 & 0.3921 & 0.2560 & 0.4473 \\
& PISA
& 0.2016 & 0.2821 & 0.2222 & 0.3454 & 0.2356 & 0.3983
& 0.2680 & 0.3586 & 0.2859 & 0.4140 & 0.2982 & 0.4624
& 0.2585 & 0.3352 & 0.2763 & 0.3895 & 0.2910 & 0.4476 \\
& Reloop2
& 0.2017 & 0.2963 & 0.2245 & 0.3664 & 0.2340 & 0.4035
& 0.2573 & 0.3496 & 0.2797 & 0.4185 & 0.2921 & 0.4669
& 0.2733 & 0.3584 & 0.2909 & 0.4125 & 0.3013 & \underline{0.4534} \\
& CCD
& 0.2024 & 0.2853 & 0.2222 & 0.3464 & \underline{0.2360} & 0.4006
& 0.2733 & 0.3633 & 0.2906 & 0.4167 & 0.3030 & 0.4660
& 0.2524 & 0.3345 & 0.2707 & 0.3909 & 0.2837 & 0.4424 \\
& LLMD4Rec
& \underline{0.2145} & \underline{0.3183} & \underline{0.2316} & \underline{0.3702} & 0.2386 & 0.3980
& \underline{0.2875} & \underline{0.3751} & \underline{0.3050} & \underline{0.4290} & \underline{0.3159} & 0.4720
& \underline{0.2902} & \underline{0.3684} & \underline{0.3049} & \underline{0.4132} & \underline{0.3123} & 0.4419 \\
& CoT-Rec
& 0.1950 & 0.2954 & 0.2186 & 0.3677 & 0.2307 & \underline{0.4154}
& 0.2002 & 0.3069 & 0.2350 & 0.4138 & 0.2518 & \textbf{0.4798}
& 0.1996 & 0.3064 & 0.2294 & 0.3979 & 0.2447 & \textbf{0.4572} \\
& \cellcolor[HTML]{D9D9D9}SCoRD
& \cellcolor[HTML]{D9D9D9}\textbf{0.3389*}
& \cellcolor[HTML]{D9D9D9}\textbf{0.3731*}
& \cellcolor[HTML]{D9D9D9}\textbf{0.3464*}
& \cellcolor[HTML]{D9D9D9}\textbf{0.3961*}
& \cellcolor[HTML]{D9D9D9}\textbf{0.3532*}
& \cellcolor[HTML]{D9D9D9}\textbf{0.4235*}
& \cellcolor[HTML]{D9D9D9}\textbf{0.4068*}
& \cellcolor[HTML]{D9D9D9}\textbf{0.4460*}
& \cellcolor[HTML]{D9D9D9}\textbf{0.4105*}
& \cellcolor[HTML]{D9D9D9}\textbf{0.4575*}
& \cellcolor[HTML]{D9D9D9}\textbf{0.4150*}
& \cellcolor[HTML]{D9D9D9}\underline{0.4756}
& \cellcolor[HTML]{D9D9D9}\textbf{0.3885*}
& \cellcolor[HTML]{D9D9D9}\textbf{0.4195*}
& \cellcolor[HTML]{D9D9D9}\textbf{0.3927*}
& \cellcolor[HTML]{D9D9D9}\textbf{0.4321*}
& \cellcolor[HTML]{D9D9D9}\textbf{0.3955*}
& \cellcolor[HTML]{D9D9D9}0.4431 \\

 \hline \hline
 \multicolumn{2}{c|}{\multirow{2}{*}{\textbf{Movies}}} & \multicolumn{6}{c|}{\textbf{After $D_2$}} & \multicolumn{6}{c|}{\textbf{After $D_3$}} & \multicolumn{6}{c}{\textbf{After $D_4$}} \\
\multicolumn{2}{c|}{} & \textbf{N@5} & \textbf{H@5} & \textbf{N@10} & \textbf{H@10} & \textbf{N@20} & \textbf{H@20} & \textbf{N@5} & \textbf{H@5} & \textbf{N@10} & \textbf{H@10} & \textbf{N@20} & \textbf{H@20} & \textbf{N@5} & \textbf{H@5} & \textbf{N@10} & \textbf{H@10} & \textbf{N@20} & \textbf{H@20} \\
\hline\hline
\multirow{8}{*}{\rotatebox[origin=c]{90}{\textbf{Reranker}}}
& Full-Batch
& 0.5625 & 0.5871 & 0.5700 & 0.6100 & 0.5852 & 0.6717
& 0.4905 & 0.5236 & 0.5033 & 0.5639 & 0.5197 & 0.6309
& 0.4451 & 0.4808 & 0.4575 & 0.5192 & 0.4706 & 0.5725 \\
& Fine-Tune
& 0.5647 & 0.5805 & 0.5715 & 0.6018 & 0.5883 & 0.6705
& 0.5218 & 0.5392 & 0.5338 & 0.5766 & 0.5508 & 0.6464
& 0.4972 & 0.5200 & 0.5087 & 0.5553 & 0.5248 & 0.6212 \\
& PISA
& 0.5638 & 0.5781 & 0.5708 & 0.6002 & 0.5881 & 0.6713
& 0.5315 & 0.5476 & 0.5440 & 0.5865 & 0.5637 & \underline{0.6662}
& 0.5050 & 0.5239 & 0.5173 & 0.5616 & \underline{0.5383} & \underline{0.6471} \\
& Reloop2
& 0.5613 & \underline{0.5875} & 0.5698 & \underline{0.6141} & 0.5837 & 0.6713
& 0.5336 & \underline{0.5603} & 0.5423 & 0.5872 & 0.5589 & 0.6549
& 0.5036 & \underline{0.5349} & 0.5108 & 0.5576 & 0.5257 & 0.6180 \\
& CCD
& 0.5621 & 0.5756 & 0.5713 & 0.6043 & 0.5892 & \underline{0.6774}
& \underline{0.5377} & 0.5568 & \underline{0.5485} & \underline{0.5907} & \underline{0.5677} & \textbf{0.6690}
& 0.4997 & 0.5200 & 0.5115 & 0.5561 & 0.5336 & 0.6463 \\
& LLMD4Rec
& 0.5657 & 0.5814 & 0.5732 & 0.6047 & 0.5859 & 0.6570
& 0.5320 & 0.5533 & 0.5394 & 0.5766 & 0.5554 & 0.6415
& \underline{0.5068} & 0.5310 & \underline{0.5178} & \underline{0.5655} & 0.5333 & 0.6290 \\
& CoT-Rec
& \underline{0.5658} & 0.5846 & \underline{0.5737} & 0.6096 & \underline{0.5903} & \textbf{0.6783}
& 0.5354 & 0.5540 & 0.5437 & 0.5801 & 0.5643 & 0.6648
& 0.4990 & 0.5176 & 0.5097 & 0.5514 & 0.5347 & \textbf{0.6541} \\
& \cellcolor[HTML]{D9D9D9}SCoRD
& \cellcolor[HTML]{D9D9D9}\textbf{0.6072*}
& \cellcolor[HTML]{D9D9D9}\textbf{0.6492*}
& \cellcolor[HTML]{D9D9D9}\textbf{0.6141*}
& \cellcolor[HTML]{D9D9D9}\textbf{0.6701*}
& \cellcolor[HTML]{D9D9D9}\textbf{0.6149*}
& \cellcolor[HTML]{D9D9D9}0.6733
& \cellcolor[HTML]{D9D9D9}\textbf{0.5831*}
& \cellcolor[HTML]{D9D9D9}\textbf{0.6295*}
& \cellcolor[HTML]{D9D9D9}\textbf{0.5916*}
& \cellcolor[HTML]{D9D9D9}\textbf{0.6556*}
& \cellcolor[HTML]{D9D9D9}\textbf{0.5927*}
& \cellcolor[HTML]{D9D9D9}0.6598
& \cellcolor[HTML]{D9D9D9}\textbf{0.5658*}
& \cellcolor[HTML]{D9D9D9}\textbf{0.6188*}
& \cellcolor[HTML]{D9D9D9}\textbf{0.5720*}
& \cellcolor[HTML]{D9D9D9}\textbf{0.6376*}
& \cellcolor[HTML]{D9D9D9}\textbf{0.5733*}
& \cellcolor[HTML]{D9D9D9}0.6424 \\
\hline
\multirow{8}{*}{\rotatebox[origin=c]{90}{\textbf{Retriever}}}
& Full-Batch
& 0.3386 & 0.5041 & 0.3759 & 0.6177 & 0.3899 & 0.6721
& 0.2041 & 0.3451 & 0.2604 & 0.5194 & 0.2886 & 0.6302
& 0.1590 & 0.2839 & 0.2086 & 0.4369 & 0.2433 & 0.5733 \\
& Fine-Tune
& 0.3529 & 0.5135 & 0.3850 & 0.6112 & 0.4002 & 0.6705
& 0.2663 & 0.3980 & 0.3130 & 0.5413 & 0.3398 & 0.6464
& 0.2098 & 0.3310 & 0.2624 & 0.4941 & 0.2948 & 0.6212 \\
& PISA
& 0.3509 & 0.4971 & 0.3850 & 0.6018 & 0.4027 & 0.6713
& 0.3309 & 0.4686 & 0.3692 & 0.5872 & 0.3895 & 0.6662
& \underline{0.3315} & \underline{0.4792} & \underline{0.3670} & \underline{0.5875} & \underline{0.3821} & \underline{0.6471} \\
& Reloop2
& \underline{0.4168} & \underline{0.5597} & \underline{0.4411} & \underline{0.6349} & 0.4505 & 0.6713
& 0.3212 & 0.4686 & 0.3579 & 0.5822 & 0.3767 & 0.6549
& 0.2056 & 0.3224 & 0.2550 & 0.4753 & 0.2914 & 0.6180 \\
& CCD
& 0.3584 & 0.5135 & 0.3914 & 0.6149 & 0.4075 & \underline{0.6774}
& \underline{0.3773} & \underline{0.5138} & \underline{0.4076} & \underline{0.6062} & \underline{0.4237} & \textbf{0.6690}
& 0.2784 & 0.4235 & 0.3194 & 0.5498 & 0.3441 & 0.6463 \\
& LLMD4Rec
& 0.3804 & 0.5180 & 0.4076 & 0.6018 & 0.4217 & 0.6570
& 0.3386 & 0.4601 & 0.3769 & 0.5773 & 0.3934 & 0.6415
& 0.2854 & 0.4039 & 0.3239 & 0.5239 & 0.3507 & 0.6290 \\
& CoT-Rec
& 0.4146 & 0.5462 & 0.4407 & 0.6267 & \underline{0.4538} & \textbf{0.6778}
& 0.3188 & 0.4651 & 0.3554 & 0.5773 & 0.3778 & \underline{0.6648}
& 0.3142 & 0.4353 & 0.3527 & 0.5545 & 0.3781 & \textbf{0.6541} \\
& \cellcolor[HTML]{D9D9D9}SCoRD
& \cellcolor[HTML]{D9D9D9}\textbf{0.6003*}
& \cellcolor[HTML]{D9D9D9}\textbf{0.6533*}
& \cellcolor[HTML]{D9D9D9}\textbf{0.6051*}
& \cellcolor[HTML]{D9D9D9}\textbf{0.6680*}
& \cellcolor[HTML]{D9D9D9}\textbf{0.6064*}
& \cellcolor[HTML]{D9D9D9}0.6733
& \cellcolor[HTML]{D9D9D9}\textbf{0.5716*}
& \cellcolor[HTML]{D9D9D9}\textbf{0.6316*}
& \cellcolor[HTML]{D9D9D9}\textbf{0.5782*}
& \cellcolor[HTML]{D9D9D9}\textbf{0.6514*}
& \cellcolor[HTML]{D9D9D9}\textbf{0.5804*}
& \cellcolor[HTML]{D9D9D9}0.6598
& \cellcolor[HTML]{D9D9D9}\textbf{0.5572*}
& \cellcolor[HTML]{D9D9D9}\textbf{0.6149*}
& \cellcolor[HTML]{D9D9D9}\textbf{0.5634*}
& \cellcolor[HTML]{D9D9D9}\textbf{0.6337*}
& \cellcolor[HTML]{D9D9D9}\textbf{0.5656*}
& \cellcolor[HTML]{D9D9D9}0.6424 \\
 \hline
\end{tabular}
}
\vspace{-0.4cm}
\end{table*}

%% file: tables/sp_balance.tex
\begin{table}[t]
\centering
\caption{Stability (RA) and plasticity (LA) comparison.}
\small
\setlength{\tabcolsep}{10pt}
\renewcommand{\arraystretch}{0.8}
\begin{tabular}{l | l |c c c c c}
\toprule
\textbf{Dataset} & \textbf{Model} & \textbf{RA} & \textbf{LA} & \textbf{H-mean} \\
\midrule\midrule
\multirow{3}{*}{\textbf{Book}}
& CCD & 0.1159 & 0.4293 & 0.1825 \\
& PISA & 0.1037 & 0.4322 & 0.1673 \\
& SCoRD & \textbf{0.1276} & \textbf{0.4762} & \textbf{0.2013} \\
\midrule
\multirow{3}{*}{\textbf{Yelp}}
& CCD & 0.0289 & 0.3376 & 0.0532 \\
& PISA & 0.0274 & 0.3376 & 0.0507 \\
& SCoRD & \textbf{0.0478} & \textbf{0.3642} & \textbf{0.0845} \\
\midrule
\multirow{3}{*}{\textbf{Movies}}
& CCD & 0.0857 & 0.5554 & 0.1485 \\
& PISA & 0.0839 & 0.5544 & 0.1457 \\
& SCoRD & \textbf{0.1189} & \textbf{0.6034} & \textbf{0.1987} \\
\bottomrule
\end{tabular}
\label{tab:sp_balance}
\vspace{-0.1cm}
\end{table}

%% file: tables/acc_eff.tex
\begin{table}[t]
\centering
\caption{Accuracy-efficiency trade-off comparison.}
\small
\renewcommand{\arraystretch}{0.8}
\resizebox{\linewidth}{!}{
\begin{tabular}{l | l |c c c c c c c}
\toprule
\textbf{Dataset} & \textbf{Model} & \textbf{Gen time} & \textbf{Tr. time} & \textbf{Inf. time} & \textbf{N@5} & \textbf{H@5}\\
\midrule\midrule
\multirow{3}{*}{\textbf{Book}}
& LLMD4Rec & N/A & 157m & 75s & 0.4740 & 0.5067 \\
& CoT-Rec & 15.4h & 32.4m & 75s & 0.4682 & 0.5132 \\
& SCoRD & 1.4h & 106.5m & 75s & \textbf{0.5088} & \textbf{0.5846} \\
\midrule
\multirow{3}{*}{\textbf{Yelp}}
& LLMD4Rec & N/A & 156m & 89s & 0.3804 & 0.3925 \\
& CoT-Rec & 17.6h & 37m & 88s & 0.3649 & 0.3934 \\
& SCoRD & 1.7h & 125.3m & 88s & \textbf{0.3866} & \textbf{0.4136} \\
\midrule
\multirow{3}{*}{\textbf{Movies}}
& LLMD4Rec & N/A & 78m & 43s & 0.5348 & 0.5552 \\
& CoT-Rec & 5.4h & 19.5m & 43s & 0.5334 & 0.5521 \\
& SCoRD & 0.5h & 52m & 44s & \textbf{0.5719} & \textbf{0.6266} \\
\bottomrule
\end{tabular}
}
\vspace{-0.1cm}
\label{tab:acc_eff}
\end{table}

%% file: tables/ablation_avg.tex
\begin{table}[t]
\centering
\caption{Ablation study across three stages.}
\small
\resizebox{\columnwidth}{!}{
\setlength{\tabcolsep}{5pt}
\renewcommand{\arraystretch}{0.95}
\begin{tabular}{l|cc|cc}
\toprule
\multicolumn{1}{c|}{\multirow{2}{*}[0.1em]{\textbf{Model Variant}}}
& \multicolumn{2}{c|}{\textbf{Reranker}}
& \multicolumn{2}{c}{\textbf{Retriever}} \\
& N@5 & H@5 & N@5 & H@5 \\
\midrule\midrule
\textbf{Stage 1} & & & & \\
\hspace{2mm} w/o $\mathcal{A}$ {\small(removes both $\mathcal{L}_{\text{intent}}$ and $\tilde{\mathbf{h}}_u$)}
& 0.3748 & 0.4945 & 0.2675 & 0.3972 \\
\hspace{2mm} w/o $\mathcal{L}_{\text{KD}}$ {}
& 0.4959 & 0.5748 & 0.4935 & 0.5798 \\
\hspace{2mm} w/o $\mathcal{A}$ \& $\mathcal{L}_{\text{KD}}$
& 0.3598 & 0.4809 & 0.2467 & 0.3768 \\
& & & & \\[-5pt]
\textbf{Stage 2} & & & & \\
\hspace{2mm} w/o $\mathcal{L}_{\text{co-intent}}$ {}
& 0.5012 & 0.5782 & 0.4960 & 0.5780 \\
\hspace{2mm} w/o $\tilde{h}_u$  {\small(uses $h_u$ instead)}
& 0.4934 & 0.5659 & 0.4790 & 0.5726 \\
\hspace{2mm} w/o Intent-drift $\mathcal{N}_u^-$ {\small(uses random $\mathcal{N}_u^-$)}
& 0.3549 & 0.4797 & 0.2455 & 0.3738 \\
\hspace{2mm} w/o $\mathcal{L}_{\text{reg}}$ {}
& 0.5061 & 0.5729 & \textbf{0.5096} & 0.5803 \\
& & & & \\[-5pt]
\textbf{Stage 3} & & & & \\
\hspace{2mm} w/o $\mathcal{L}_{\text{drift}}$ {}
& 0.4957 & 0.5746 & 0.4947 & 0.5795 \\
\hline
\cellcolor[HTML]{D9D9D9}\textbf{Full method} & \cellcolor[HTML]{D9D9D9}\textbf{0.5088} & \cellcolor[HTML]{D9D9D9}\textbf{0.5846} & \cellcolor[HTML]{D9D9D9}0.4993 & \cellcolor[HTML]{D9D9D9}\textbf{0.5852}  \\
\specialrule{0.5pt}{0pt}{0pt}
\hline
\end{tabular}
}
\label{tab:ablation_avg}
\end{table}

%% file: sections/060conclusion.tex
\section{Conclusion}
We propose \proposed, a semantic-assisted continual KD framework that enables co-adaptation of an ID-based retriever and an LLM reranker under a non-stationary data stream.
With the semantic reasoning assistant, \proposed converts the LLM's intent inference ability into reusable guidance for retriever updates. 
It supports selective reranker-to-retriever distillation, intent-guided retriever updates, and retriever-informed reranker updates. 
Extensive experiments show that \proposed improves both the retriever and the reranker, highlighting the effectiveness of semantic-assisted co-adaptation.
We expect this work to contribute to the practical deployment of LLM-based reranking pipelines in dynamic recommendation environments.